\documentclass[aip,jcp,prereprint,superscriptaddress,groupedaddress,frontmatterverbose,]{revtex4-1}

\usepackage{graphicx}
\usepackage{amsmath}
\usepackage{amssymb}
\usepackage{epstopdf}
\usepackage{color}
\usepackage{bm}
\usepackage{dcolumn}

\begin{document}

\preprint{APS/123-QED}

\title{Analysis of 2D THz-Raman spectroscopy using a non-Markovian
  Brownian oscillator model with nonlinear system-bath interactions}

\author{Tatsushi Ikeda}     \email{t.ikeda@kuchem.kyoto-u.ac.jp}
\author{Hironobu Ito}       \email{h.ito@kuchem.kyoto-u.ac.jp}
\author{Yoshitaka Tanimura} \email{tanimura@kuchem.kyoto-u.ac.jp}
\affiliation{Department of Chemistry, Graduate School of Science, Kyoto University, Kyoto 606-8502, Japan}

\date{\today}

\begin{abstract}
We explore and describe the roles of inter-molecular vibrations
employing a Brownian oscillator (BO) model with linear-linear (LL) and
square-linear (SL) system-bath interactions, which we use to analyze
two-dimensional (2D) THz-Raman spectra obtained by means of molecular
dynamics (MD) simulations. In addition to linear absorption (1D IR),
we calculated 2D Raman-THz-THz, THz-Raman-THz, and THz-THz-Raman
signals for liquid formamide, water, and methanol using an equilibrium
non-equilibrium hybrid MD simulation. The calculated 1D IR and 2D
THz-Raman signals are compared with results obtained from the LL+SL BO
model applied through use of hierarchal Fokker-Planck equations with
non-perturbative and non-Markovian noise. We find that all of the
qualitative features of the 2D profiles of the signals obtained from
the MD simulations are reproduced with the LL+SL BO model, indicating
that this model captures the essential features of the inter-molecular
motion. We analyze the fitted 2D profiles in terms of anharmonicity,
nonlinear polarizability, and dephasing time. The origins of the echo
peaks of the librational motion and the elongated peaks parallel to
the probe direction are elucidated using optical Liouville paths.
\end{abstract}

\pacs{Valid PACS appear here}
\maketitle

\section{Introduction}
\label{sec:introduction}
Molecular vibrations in condensed phases play an essential role in
various dynamic processes, including inter- and intra-molecular
couplings, and solvent dynamics, all of which entail energy exchange
as well as thermal excitations and relaxations.\cite{Mukamel-Book}
Multidimensional vibrational spectroscopy techniques make it possible
to experimentally distinguish such processes due to the sensitivity of
the nonlinear response functions utilized in these techniques to
complex dynamics.\cite{Cundiff-PT-2013-7, Mukamel-ACR-2009-42} For
intra-molecular vibrations, the roles of relaxation and dephasing are
well understood both theoretically and experimentally due to the
advent of infrared (IR) laser technologies. Methods of analysis with
theoretical models that utilize molecular dynamic simulations have
also been developed to elucidate multidimensional IR
signals.\cite{Asbury-JPCA-2004-108, Auer-PNAS-2007-104} Because the
primary inter-molecular modes, which are the objects of study in 2D IR
spectroscopy, can be separated from the other modes, as in the case of
the OH stretching mode in liquid water, stochastic models whose
parameters are obtained from classical molecular dynamics (MD)
simulations have been useful for analysis of the inter-molecular
vibrational modes. For inter-molecular vibrational modes,
two-dimensional Raman spectroscopy\cite{Tanimura-JCP-1993-99} was for
a long time the only two-dimensional spectroscopy that could be used
for experimental study. However, due to technical difficulties, such
investigations have been carried out only for
$\mathrm{CS}_{2}$,\cite{Kaufman-PRL-2002-88, Kubarych-IRPC-2003-22,
  Kubarych-JCP-2002-116, Kubarych-CPL-2003-369}
Benzene,\cite{Milne-JPCB-2006-40} and formamide\cite{Li-JCP-2008-128}
liquids. Theoretical investigations have also been limited, due to the
availability of experimental data and limitations on computational
power for simulations.

Two-dimensional THz-Raman spectroscopy, which has been studied both
theoretically\cite{Hamm-JCP-2012-136, Hamm-JCP-2012-136-note,
  Hamm-JCP-2014-141, Ito-JCP-2014-141} and
experimentally,\cite{Savolainen-PNAS-2013-110} has created a new
possibility for investigating the details of inter-molecular
vibrations.  In 2D Raman spectroscopy, the observable is defined in
terms of the three-body response function for the polarizability of
the system, $\hat{\bm{\Pi }}$ as
$R_{\mathrm{RRR}}^{(5)}(t_{2},t_{1})=-\langle [[\hat{\bm{\Pi }}(t_{2}+t_{1}),\hat{\bm{\Pi }}(t_{1})],\hat{\bm{\Pi }}(0)]\rangle /\hbar ^{2}$,
where $\langle \dots \rangle $ represents the thermal average and
$\hat{A}(t)\equiv e^{i\hat{H}_{\mathrm{S}}t/\hbar }\hat{A}e^{-i\hat{H}_{\mathrm{S}}t/\hbar }$ is the Heisenberg operator
for an arbitrary operator $\hat{A}$.\cite{Tanimura-JCP-1993-99} In the case
of 2D THz-Raman spectroscopy, the response function consists of one
polarizability, $\hat{\bm{\Pi }}$, and two dipole operators, $\hat{\bm{\mu }}$, and there
are three different measurements, which depend upon the sequence of
the Raman and THz pulses as
$R_{\mathrm{RTT}}^{(3)}(t_{2},t_{1})=-\langle [[\hat{\bm{\mu }}(pt_{2}+t_{1}),\hat{\bm{\mu }}(t_{1})],\hat{\bm{\Pi }}(0)]\rangle /\hbar ^{2}$,
$R_{\mathrm{TRT}}^{(3)}(t_{2},t_{1})=-\langle [[\hat{\bm{\mu }}(t_{2}+t_{1}),\hat{\bm{\Pi }}(t_{1})],\hat{\bm{\mu }}(0)]\rangle /\hbar ^{2}$,
and
$R_{\mathrm{TTR}}^{(3)}(t_{2},t_{1})=-\langle [[\hat{\bm{\Pi }}(t_{2}+t_{1}),\hat{\bm{\mu }}(t_{1})],\hat{\bm{\mu }}(0)]\rangle /\hbar ^{2}$. While
inter-molecular vibrational modes are usually both Raman and IR
active, the types of information that we can obtain from the 2D Raman
signal and each of three THz-Raman signals are different, due to the
role of the nonlinear polarizability.  Because each of the
above-mentioned response functions is defined in terms of the
three-body correlation function, the signal will vanish if the system
is harmonic and if the total dipole moment and polarizability are
linear functions of the collective coordinate, $\hat{\bm{q}}$, representing the
inter-molecular vibration, because there is an odd number of Gaussian
integrals involved in the response function:
$\mathrm{Tr}\{\hat{\bm{q}}(t_{2}+t_{1})\hat{\bm{q}}(t_{1})\hat{\bm{q}}(0)\exp (-\beta \hat{H}_{\mathrm{S}})\}$. The dipole moment is
approximated reasonably well as a linear function of $\hat{\bm{q}}$ as
$\hat{\bm{\mu }}(\hat{\bm{q}})=\mu _{1}\hat{\bm{q}} $, because the total dipole moment is a linear
function of the distance between the charges in the system, and the
nonlinear dipole-induced dipole interactions are weak. However, the
contribution of the non-linear polarization is not negligible, because
the polarizability originates in the electronic states of molecules,
which depend on the complex configurations of the atoms and
molecules. For this reason the polarizability is expressed in a Taylor
expansion form as $\hat{\bm{\Pi }}(\hat{\bm{q}})=\Pi _{1}\hat{\bm{q}}+\Pi _{2}\hat{\bm{q}}^{2}/2$.  Because $\hat{\bm{\Pi }}$ has
this non-linear form, the three response functions give above,
representing the observables in 2D THz-Raman spectroscopy experiments,
provide information about three different physical
processes. \cite{Hamm-JCP-2012-136-note} Contrastingly, because 2D
Raman spectroscopy experiments measure just a single observables, they
do not provide such a detailed picture of the physical system. The
richness of the information obtained through 2D THz-Raman spectroscopy
allows for a detailed analysis of inter-molecular vibrational modes.
Note that the optical setup for the TTR measurement differs
significantly from those for the RTT and TRT measurements. This is
because the RTT and TRT responses are detected as the emission of THz
signals, while the TTR response is detected as an induced Raman
signal.

Although we can obtain relatively reliable 2D THz-Raman signals using
the full MD simulation techniques developed for 2D Raman
spectroscopy,\cite{Ma-PRL-2000-85, Saito-PRL-2002-88,
  Saito-JCP-2003-119, Saito-JCP-2006-125, Nagata-JCP-2007-126,
  Jansen-JCP-2001-114, Jansen-JCP-2003-63, Hasegawa-JCP-2006-125}
analysis of the spectra is not straightforward, due to the complexity
of the 2D profiles of the signals, which arises from the complexity of
the inter-molecular vibrational modes. As demonstrated by 2D Raman and
2D IR spectroscopy studies, a model-based analysis is useful for
treating this problem, because the 2D profile of the signal is so
sensitive to the underlying dynamics that the complex 2D profile
cannot be reproduced without capturing the essential features of the
vibrational modes.\cite{Tanimura-ACR-2009-42} While stochastic models,
which can be regarded as Brownian models with non-linear system-bath
interactions,\cite{Ishizaki-JCP-2006-125, Ishizaki-JPC-2007-111} are
recognized as versatile models for analyzing intra-molecular modes
observed in 2D IR spectroscopy experiments, it is not clear if such
models are useful in the 2D THz-Raman case. This is because in
contrast to the intra-molecular modes, which are clearly definable
with the normal mode picture, the inter-molecular modes are not
localized and change in time due to changes in the configuration of
the system molecules.

In this paper, we explore the possibility of characterizing
inter-molecular modes using a Brownian model with linear-linear (LL)
and square-liner (SL) interactions utilizing 2D THz-Raman signals
obtained from MD simulations. In order to treat a non-perturbative,
non-Markovian, and nonlinear system-bath interaction, which is
necessary to describe the effects of homogeneous and inhomogeneous
broadening in a unified manner, we employ the hierarchal equations of
motion approach.\cite{Tanimura-JPSJ-1989-58, Tanimura-PRA-1991-43,
  Tanimura-JCP-1992-96, Steffen-JPSJ-2000-69, Tanimura-JPSJ-2000-69,
  Kato-JCP-2002-117, Kato-JCP-2004-120, Tanimura-JPSJ-2006-75,
  Tanimura-JCP-accepted} The properties of inter-molecular motion are
investigated using the fitted model.

This paper is organized as follows. In
Sec. \ref{sec:full_md_simulation}, we explain the methodology for
calculating 2D THz-Raman signals from full MD simulations. In
Sec. \ref{sec:model_calculation}, we present the LL+SL BO model and
the hierarchal equations of motion formalism.  We then show how this
formalism can be used to calculate 2D signals. The MD and fitted
results obtained from the LL+SL BO model are presented and analyzed in
Sec. \ref{sec:results_and_discussion}. Section \ref{sec:conclusion} is
devoted to concluding remarks.

\section{Full MD simulation}
\label{sec:full_md_simulation}
While, to this time, the experimentally obtained 2D THz-Raman signals
are limited to the case of liquid water, in this paper we analyze 2D
signals obtained from full MD simulations for formamide, water, and
methanol. We chose these liquids from among many substances that have
been investigated in full MD studies of the 2D Raman and 2D THz-Raman
spectroscopy as characteristic examples of 2D THz-Raman signals. The
MD simulation results used in the present study of these molecules for
2D Raman-THz-THz (RTT) and THz-Raman-THz (TRT) signals were originally
presented in a previous study.\cite{Ito-JCP-2014-141} Nevertheless,
here we repeated the full MD simulations in order to also obtain
THz-THz-Raman (TTR) and infrared absorption signals, in addition to
the RTT and TRT signals. Moreover, we employed the Ewald sum for the
evaluation of the dipole and polarizability, in addition to the force
fields. This contrasts with the situation in previous studies, in
which only force fields were computed with the Ewald sum. The change
in the resulting signals due to the use of the Ewald sum in the
computation of the dipole and polarizability, however, are small.

\subsection{Models and simulation details}
Based on the MD simulations, we calculated the linear absorption
(1DIR) spectrum and 2D THz-Raman signals of liquid formamide, water,
and methanol. Each system consisted of 108 molecules in a cubic box
with periodic boundary conditions. The interactions between the
molecules were modeled by a modified T
potential,\cite{Sagarik-JCP-1987-86, Wojcik-JCP-2000-113} the
TIP4P/2005 potential,\cite{Abascal-JCP-2005-23} and the B3
potential\cite{Walser-JCP-2000-112} for formamide, water, and
methanol, respectively.

The interaction potentials were cut off smoothly at a distance equal
to a half the length of the system using a switching function, and the
long-range Coulomb interactions were calculated with the Ewald
sum. The intra-molecular geometries were kept rigid throughout the
simulations, using a constraint provided by the RATTLE algorithm. The
equations of motion were integrated using the velocity-Verlet
algorithm with time steps of $5.0$ $\mathrm{fs}$ for formamide and $2.5$ $\mathrm{fs}$
for water and methanol.  The system volume and total energy were fixed
after the completion of the isothermal simulations carried out for
equilibration. The conditions of the simulation were set such that the
average densities were $1.120$ $\mathrm{g}/\mathrm{cm}^{3}$ for formamide, $0.997$ $\mathrm{g}/\mathrm{cm}^{3}$
for water, and $0.786$ $\mathrm{g}/\mathrm{cm}^{3}$ for methanol. The temperature was set
to $300$ $\mathrm{K}$.  The permanent molecular polarizability of each liquid
was utilized with the atomic polarizability for formamide and
methanol,\cite{Applequist-JACS-1972-94} and the Huiszoon
polarizability for water.\cite{Huiszoon-MP-1986-58}

\subsection{Molecular polarizability and dipole}
While the MD simulations were carried out using the permanent
polarizability, we calculated the 2D THz-Raman signals using a
full-order dipole-induced-dipole (DID) polarizability model. We did
this because the 2D profiles are extremely sensitive to the accuracy
of the calculated optical observables. In the DID polarizability
model, the expression determining the polarizability of a molecule
includes contributions from other molecules, and interactions between
molecules are defined with respect to the centers of individual
molecules. \cite{Ladanyi-CPL-1985-121, Applequist-JACS-1972-94} The
total polarizability of the system in a MD simulation is given by
$\bm{\Pi }(t)=\sum _{i}\bm{\Pi }_{i}$, where $\bm{\Pi }_{i}$ is the polarizability of the
$i$th molecule, expressed as
\begin{align}
  \bm{\Pi }_{i}=\bm{\alpha }_{i}-\sum _{j\neq i}\bm{\alpha }_{i}\bm{T}_{ij}\bm{\Pi }_{j}.
\end{align}
In this expression, $\bm{\alpha }_{i}$ is the permanent molecular
polarizability of the $i$th molecule in isolation in the laboratory
frame, and $\bm{T}_{ij}$ is the dipole-dipole interaction tensor
\begin{align}
  \bm{T}_{ij}=\frac{\bm{1}}{r^{3}_{ij}}-3\frac{\bm{r}_{ij}\otimes \bm{r}_{ij}}{r^{5}_{ij}}.
\end{align}
Here, $\bm{r}_{ij}$ is the vector from the center of mass of molecule $i$ to
the center of mass of molecule $j$, and $r_{ij}=|\bm{r}_{ij}|$. Also, $\bm{1}$ and
$\otimes $ are the unit matrix and the tensor product, respectively. In order
to properly take into account the effect of the long-range interaction
on the molecular polarizability, we employ the Ewald sum for
$\bm{T}_{ij}$. This effect has been ignored in previous MD simulations of 2D
Raman and 2D THz-Raman spectroscopy systems. However, the contribution
of this effect in the 2D THz-Raman case is minor in comparison with
that in the 2D Raman case.

The total dipole moment is evaluated as
$\bm{\mu }(t)=\sum _{i}\bm{\mu }_{i}^{\mathrm{perm}}+\sum _{i}\bm{\mu }_{i}^{\mathrm{ind}}$, where $\bm{\mu }_{i}^{\mathrm{perm}}$ and
$\bm{\mu }_{i}^{\mathrm{ind}}$ are the permanent and induced molecular dipole moments of
molecule $i$, respectively. The induced molecular dipole moment is
expressed in terms of the interaction tensor as
\begin{align}
  \bm{\mu }_{i}^{\mathrm{ind}}=\bm{\mu }_{i}\left(\bm{E}_{i}^{\mathrm{perm}}-\sum _{j\neq i}\bm{T}_{ij} \bm{\mu }_{j}^{\mathrm{ind}}\right),
\end{align}
where $\bm{E}_{i}^{\mathrm{perm}}$ is the electrostatic field at molecule $i$ created by
all the other molecules in the system. This is evaluated as
$\bm{E}_{i}^{\mathrm{perm}}=\sum _{j\neq i}\sum _{l}q^{\mathrm{perm}}_{l_j}\bm{r}_{i{l_j}}/r^{3}_{i{l_j}}$, where $\bm{r}_{i{l_j}}$
is the vector between the center of mass of molecule $i$ and that of
atom $l$ in molecule $j$ .

\subsection{One- and two-dimensional signals}
It should be noted that the majority of MD simulations of 2D IR
spectroscopy systems performed to this time have been carried out to
obtain the parameter values for stochastic models. Full MD simulations
have been carried out mostly in the cases of low frequency vibrational
modes.\cite{Hasegawa-JCP-2008-128, Yagasaki-JCP-2008-128,
  Jeon-JPB-2014-118} This is because the primary inter-molecular
modes, which are the objects of study in 2D IR spectroscopy, can be
separated from the other modes rather easily, as in the case of the OH
stretching mode in liquid water. Contrastingly in the 2D THz-Raman
case, it is not easy to find primary modes, because the objects of
study in this cases are inter-molecular vibrations that depend on the
complicated nature of molecular ensembles, whose configurations change
in time. Thus, we have to evaluate 2D signals directly from the MD
simulations. Because quantum mechanical effects are minor for
low-frequency inter-molecular modes, due to their small thermal
activation energies, unlike the case of intra-molecular motion,
\cite{Sakurai-JPCA-2011-115} and because 2D THz-Raman spectroscopy
employs the three-body correlation function with two time variables,
instead of the four-body correlation function with three time
variables employed in 2D IR spectroscopy, the full MD simulation
approach is practical. For this reason with our approach, we were able
to carry out full MD simulations to directly evaluate 2D signals.

Although our MD and model calculations are fully classical, we start
from the quantum expressions for the response functions, because their
classical expressions in the MD and model calculations are most easily
derived by taking the classical limit of the quantum expressions.  The
optical observables in 1D and 2D spectroscopies are represented
respectively expressed by two- and three-body response functions of
the forms\cite{Tanimura-JCP-1993-99, Tanimura-JPSJ-2006-75}
\begin{align}
  R(t)=\frac{i}{\hbar }\langle [\hat{\bm{A}}(t),\hat{\bm{B}}(0)]\rangle 
  \label{eq:AB}
\end{align}
and
\begin{align}
  R(t_{2},t_{1})=\left(\frac{i}{\hbar }\right)^{2}\langle [[\hat{\bm{A}}(t_{2}+t_{1}),\hat{\bm{B}}(t_{1})],\hat{\bm{C}}(0)]\rangle ,
  \label{eq:ABC}
\end{align}
where $\hat{\bm{A}}$, $\hat{\bm{B}}$, and $\hat{\bm{C}}$ can be the total dipole moment,
$\hat{\bm{\mu }}$, or the total polarizability of the molecules, $\hat{\bm{\Pi }}$.
For low-frequency inter-molecular vibrations, we can take the
classical limit, $\hbar \rightarrow 0$.  The commutator and operators are then replaced
by the Poisson bracket and c-number observables as
\begin{align}
  -\frac{i}{\hbar }[\hat{\bm{A}},\hat{\bm{B}}]\rightarrow \{\bm{A},\bm{B}\}_{\mathrm{PB}}\equiv \frac{\partial \bm{A}}{\partial \bm{q}}\frac{\partial \bm{B}}{\partial \bm{p}}-\frac{\partial \bm{A}}{\partial \bm{p}}\frac{\partial \bm{B}}{\partial \bm{q}} \label{eq:pb}.
\end{align}
Using $\{e^{-\beta H_{0}(\bm{p},\bm{q})},{\bm{A}}(t)\}_{\mathrm{PB}}=\beta e^{-\beta H_{0}(\bm{p},\bm{q})}\dot{\bm{A}}(t)$ for a
molecular Hamiltonian $H_{0}(\bm{p},\bm{q})$, we obtain the expression for the<
linear response function, for example, for 1D IR as
\begin{align}
  R^{(1)}_{\mathrm{IR}}(t)=\beta \langle \bm{\mu }_{\mathrm{eq}}(t)\dot{\bm{\mu }}_{\mathrm{eq}}(0)\rangle 
\end{align}
and 
\begin{align}
  I_{\mathrm{IR}}^{(1)}(\omega )\propto  \omega \mathrm{Im}\int _{0}^{\infty }\mathrm{d}te^{i\omega t}R_{\mathrm{IR}}^{(1)}(t).
\label{IIR}
\end{align}
It should be noted that the quantum correction factor,
$\tanh (\beta \hbar \omega /2)$, is usually included in Eq. \eqref{IIR}
to allow comparison with experimentally obtained IR signals, but here
we do not include it.  Instead, as the classical limit, we only
multiply by $\omega$. \cite{Berens-JCP-1981-74} We can evaluate the
above quantities easily by calculating $\bm{\mu }_{\mathrm{eq}}(\bm{q}(t))$ from samples of
molecular trajectories $\bm{q}(t)$ that are obtained from the equilibrium
MD simulation.

For the 2D case, the response function in the classical limit is
expressed as\cite{Ma-PRL-2000-85, Saito-PRL-2002-88}
\begin{align}
  R(t_{2},t_{1})=\langle \{\{\bm{A }(t_{2}),\bm{B }(0)\}_{\mathrm{PB}},\bm{C }(-t_{1})\}_{\mathrm{PB}}\rangle
  \label{eq:2D_response}
\end{align}
As in the 1D case, we can calculate the above response function using
$\bm{\mu }(t)$ and $\bm{\Pi }(t)$ evaluated from the molecular trajectories
$\bm{p}(t)$ and $\bm{q}(t)$ obtained from the equilibrium MD
simulations. \cite{Ma-PRL-2000-85, Saito-PRL-2002-88,
  Saito-JCP-2003-119, Saito-JCP-2006-125, Nagata-JCP-2007-126} The
convergence of the signal is, however, very slow due to the effect of
the stability matrix element in the double Poisson brackets. However,
there is different approach, the non-equilibrium finite field
approach, that does not have this convergence problem. In this
approach, the double Poisson brackets are evaluated as
$-\{\bm{A}(t'),\bm{B}(t)\}_{\mathrm{PB}}=(\bm{A}_{+\bm{B}(t)}(t')-\bm{A}_{-\bm{B}(t)}(t'))/2F$, where $\bm{A}_{\pm \bm{B}(t)}(t')$ is
the observable corresponding to $\bm{A}(t')$ calculated from the
trajectories subjected to the weak perturbations
$\pm (-F\delta (\tau -t)\bm{B}(\tau ))$, with the electric field $\pm F$
acting on $\bm{B}(\tau )$.\cite{Jansen-JCP-2001-114, Jansen-JCP-2003-63}
But, this approach is computationally intensive, and for this reason,
here we employed a hybrid approach, which utilizes both the
equilibrium and non-equilibrium approaches in order to reduce the
computational cost further.\cite{Hasegawa-JCP-2006-125}

In our hybrid approach, we evaluate $\dot{\bm{C}}(-t_{1})\equiv \mathrm{d}\bm{C}(t)/\mathrm{d}t|_{t=-t_{1}}$
with equilibrium MD simulations, while $\bm{A}_{\pm \bm{B}(0)}(t_{2})$ with
non-equilibrium MD simulations. As a result, the hybrid expressions
for the 2D Raman-THz-THz, THz-Raman-THz, and THz-THz-Raman signals
become
\begin{align}
  R^{(3)}_{\mathrm{RTT}}(t_{2},t_{1})&=\frac{\beta }{E_{1}}\langle \left({\bm{\mu }}_{+\bm{\mu }(0)}(t_{2})-{\bm{\mu }}_{-\bm{\mu }(0)}(t_{2})\right) \dot{\bm{\Pi }}_{\mathrm{eq}}(-{t_{1}})\rangle , \\
  R^{(3)}_{\mathrm{TRT}}(t_{2},t_{1})&=\frac{2\beta }{E_{1}E_{2}}\langle \left({\bm{\mu }}_{+\bm{\Pi }(0)}(t_{2})-{\bm{\mu }}_{-\bm{\Pi }(0)}(t_{2})\right) \dot{\bm{\mu }}_{\mathrm{eq}}(-{t_{1}})\rangle , 
\end{align}
and
\begin{align}
  R^{(3)}_{\mathrm{TTR}}(t_{2},t_{1})&=\frac{\beta }{E_{1}}\langle \left({\bm{\Pi }}_{+\bm{\mu }(0)}(t_{2})-{\bm{\Pi }}_{-\bm{\mu }(0)}(t_{2})\right) \dot{\bm{\mu }}_{\mathrm{eq}}(-t_{1})\rangle ,
\end{align}
where $E_{j}$ and $\beta =1/k_{\mathrm{B}}T$ are the external electric field of the
$j$th pulse and the inverse temperature divided by the Boltzmann
constant.

\section{Model calculation}
\label{sec:model_calculation}
\subsection{Brownian oscillator model with nonlinear interaction}
In order to analyze the 2D signals of inter-vibrational modes, we
consider a model that consists of a primary oscillator mode
nonlinearly coupled to the other modes, which are regarded as a bath
system. This bath system is represented by an ensemble of harmonic
oscillators.  The primary mode may change in time or be
inhomogeneously distributed. We can describe both situations within a
unified framework by adjusting the bath parameter variables.  The
model is constructed by extending a Brownian (or Caldeira-Leggett)
Hamiltonian\cite{Caldeira-PA-1983-121, Grabert-PR-1988-115} to include
a nonlinear system-bath interaction. We write
\begin{align}
  \hat{H}&=\hat{H}_{\mathrm{S}}+\hat{H}_{\mathrm{B}}+\hat{H}_{\mathrm{I}},
  \label{eq:h_total}
\end{align}
where
\begin{align}
  \hat{H}_{\mathrm{S}}=\frac{\hat{p}^{2}}{2m}+U(\hat{q})
  \label{eq:h_system}
\end{align}
is the Hamiltonian for the system with mass $m$, momentum $\hat{p}$
and potential $U(\hat{q})$,
\begin{align}
  \hat{H}_{\mathrm{B}}=\sum _{j}\left(\frac{\hat{p}_{j}^{2}}{2m_{j}}+\frac{m_{j}\omega _{j}^{2}\hat{x}_{j}^{2}}{2} \right)+ \sum _{j}\left(\frac{\alpha _{j}^{2}V^{2}(\hat{q})}{2m_{j}\omega _{j}^{2}}\right)
  \label{eq:h_bath}
\end{align}
is the bath Hamiltonian with the momentum, coordinate, mass, and
frequency of the $j$th bath oscillator given by $\hat{p}_{j}$, $\hat{x}_{j}$, $m_{j}$ and
$\omega _{j}$, respectively, and
\begin{align}
  \hat{H}_{\mathrm{I}}&=- V(\hat{q})\sum _{j}\alpha _{j}\hat{x}_{j},
  \label{eq:h_int}
\end{align}
is the system-bath interaction, which consists of linear-linear (LL)
and square-linear (SL) system-bath interactions,
$V(\hat{q})\equiv V_{\mathrm{LL}}\hat{q}+V_{\mathrm{SL}}\hat{q}^{2}/2$, with coupling strengths $V_{\mathrm{LL}}$, $V_{\mathrm{SL}}$,
and $\alpha _{j}$.\cite{Okumura-PRE-1997-56} This model has been used to derive
predictions for 2D Raman\cite{Steffen-JPSJ-2000-69,
  Tanimura-JPSJ-2000-69, Kato-JCP-2002-117, Kato-JCP-2004-120,
  Tanimura-JPSJ-2006-75} and 2D IR
signals.\cite{Ishizaki-JCP-2006-125,
  Ishizaki-JPC-2007-111,Tanimura-ACR-2009-42} The last term of the
bath Hamiltonian is the counter-term, which maintains the
translational symmetry of the system in the case $U(q)=0$.

The sum of the bath coordinates $\hat{X}\equiv \sum _{j}\alpha _{j}\hat{x}_{j}$ acts as a
collective coordinate that modulates the
system.\cite{Tanimura-JPSJ-2006-75} As illustrated in
Ref. \onlinecite{Tanimura-ACR-2009-42}, while the LL interaction
shifts the potential, the SL interaction changes its curvature.
Although in the anharmonic potential case, the LL interaction also
changes the curvature of the potential, we can ignore this effect if
the anharmonicity is
weak.\cite{Sakurai-JPCA-2011-115,Tanimura-ACR-2009-42} Next we
introduce the spectral distribution function,
$J(\omega )\equiv \sum _{j}\alpha _{j}^{2}\hbar \delta (\omega -\omega _{j})/2m_{j}\omega _{j}$, which
characterizes the bath and system-bath coupling. We assume that $J(\omega )$
has an Ohmic form with a Lorentzian
cutoff:\cite{Tanimura-JPSJ-1989-58,Tanimura-PRA-1991-43,
  Tanimura-JCP-1992-96,Steffen-JPSJ-2000-69, Tanimura-JPSJ-2000-69,
  Kato-JCP-2002-117, Kato-JCP-2004-120,
  Tanimura-JPSJ-2006-75,Tanimura-JCP-accepted}
\begin{align}
  J(\omega )=\frac{\hbar m\zeta }{2\pi }\frac{\omega \gamma ^{2}}{\gamma ^{2}+\omega ^{2}},
\end{align}
where $\zeta $ is the system-bath coupling strength, and $\gamma $ represents
the width of the spectral distributuion.

Writing the classical collective coordinate corresponds as $X$, we
have the correlation function,
$\langle X(t)X(0)\rangle \propto e^{-\gamma \left|t\right|}$. This indicates that the bath
oscillators interact with the system in the form of Gaussian Markovian
noise with correlation time $\tau =1/\gamma $.\cite{Tanimura-JPSJ-2006-75}
\begin{figure}[tbp]
  \centering
  \includegraphics[scale=1.2]{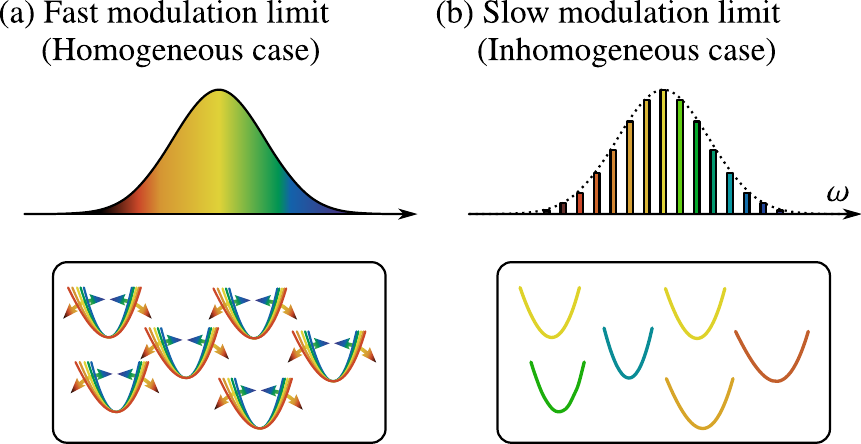}
  \caption{ Schematic illustration of the relation between line broadening
    and modulation of the potential system perturbed by the SL system-bath
    interaction. (a)The fast modulation limit corresponds to the
    homogeneous broadening case, whereas (b) the slow modulation limit
    corresponds to the inhomogeneous broadening case. }
  \label{fig:homo_inhomo}
\end{figure}
Because the SL interaction affects the frequency of the potential, the
fast modulation limit ($\tau \rightarrow 0$) of the SL interaction corresponds to the
case of a homogeneous distribution, as depicted in
Fig. \ref{fig:homo_inhomo}(a). By contrast, the slow modulation limit
($\tau \rightarrow \infty $) corresponds to the case of an inhomogeneous distribution, as
depicted in Fig. \ref{fig:homo_inhomo}(b). Note that SL or LL+SL BO
model has the ``mode mixing in polarization''\cite{Saito-JCP-108-1998}
because the modes included in collective coordinate $q$ are frequency
distributed by SL interaction and are mixed by nonlinear
polarizability $\Pi _{2}q^{2}$.

\subsection{Classical hierarchal Fokker-Planck equations}
Because we wish to explore the effects of anharmonicity, nonlinear
polarizability, vibrational dephasing, and homogeneous and
inhomogeneous broadening within a unified framework, we must employ a
kinetic equation that can treat thermal fluctuations as well as
dissipation in a non-perturbative, non-Markovian manner. The reduced
hierarchal equations of motion (HEOM) satisfy all of the requirements
mentioned above and is ideal for the present
study.\cite{Tanimura-JPSJ-1989-58, Tanimura-PRA-1991-43,
  Tanimura-JCP-1992-96, Tanimura-JCP-accepted, Kato-JCP-2002-117,
  Kato-JCP-2004-120, Tanimura-JPSJ-2006-75, Steffen-JPSJ-2000-69,
  Tanimura-JPSJ-2000-69} While we must use the quantum form of the
equations to calculate the signals for high frequency intra-molecular
modes,\cite{Sakurai-JPCA-2011-115} we can employ the classical form
for low frequency inter-molecular modes.

For the LL+SL BO Hamiltonian, given in
Eqs. \eqref{eq:h_total}-\eqref{eq:h_int}, the HEOM for the classical
distribution function are expressed
as\cite{Kato-JCP-2002-117, Kato-JCP-2004-120, Tanimura-JPSJ-2006-75}
\begin{align}
  \frac{\partial W^{(n)}(p,q;t)}{\partial t}&=-\left(\mathcal{\hat{L}}_{\mathrm{cl}}+n\gamma \right)W^{(n)}(p,q;t)-n\gamma \hat{\Theta }W^{(n-1)}(p, q; t)-\hat{\Phi }W^{(n+1)}(p,q;t)
  \label{eq:heom1}
\end{align}
for $0\leq n<N$ and
\begin{align}
  \frac{\partial W^{(N)}(p,q;t)}{\partial t}&=-\left(\mathcal{\hat{L}}_{\mathrm{cl}}+N\gamma -\hat{\Phi }\hat{\Theta }\right)W^{(N)}(p,q;t) -N\gamma \hat{\Theta }W^{(N-1)}(p, q; t)
  \label{eq:heom2}.
\end{align}
 In the HEOM approach, only the first element, $W^{(0)}(p,q;t)$, has
 physical meaning, while the other elements, $W^{(n)}(p,q;t)$
 $(1 \le n \le N)$, are introduced in the numerical calculations in order to
 treat the non-perturbative, non-Markovian system-bath interaction. We
 choose $N$ to satisfy $N \gg \omega _{\mathrm{c}}/\gamma $, where $\omega _{\mathrm{c}}$ is the
 characteristic frequency of the system.

The classical Liouvillian of the system, $\mathcal{\hat{L}}_{\mathrm{cl}}$, is defined by
\begin{align}
  \mathcal{\hat{L}}_{\mathrm{cl}}&\equiv \frac{p}{m}\frac{\partial }{\partial q}-U' (q)\frac{\partial }{\partial p}, \label{eq:liouvillian}
\end{align}
where the dash is defined as $A'(q)\equiv {\partial A(q)}/{\partial q}$ for an arbitrary
function $A(q)$.  The operators $\hat{\Phi }$ and $\hat{\Theta }$ describe the
energy exchange between the system and the heat bath for the inverse
correlation time $\gamma $. They are defined as
\begin{align}
  \hat{\Phi }&\equiv -V'(q)\frac{\partial }{\partial p}
\label{eq:phi}
\end{align}
and
\begin{align}
  \hat{\Theta }&\equiv -\zeta V'(q)\left(p+\frac{m}{\beta }\frac{\partial }{\partial p}\right) \label{eq:theta},
\end{align}
with $\beta =1/k_{\mathrm{B}}T$ and
\begin{align}
  V'(q)&\equiv V_{\mathrm{LL}}+V_{\mathrm{SL}}q.
\end{align}
The thermal equilibrium distribution, $W^{\mathrm{eq}}(p,q)$, is expressed in
terms of the HEOM elements evaluated from the steady-state solution of
the HEOM.  Note that Eqs. \eqref{eq:heom1}-\eqref{eq:heom2} reduce to
the Kramers equation in the limit $N\rightarrow 0$ with
$V_{\mathrm{SL}}=0$.\cite{Risken-Book}

Hereafter, we employ the dimensionless coordinate and momentum defined
by $\bar{q}\equiv \omega _{0}\sqrt {m\beta /2}\times q$ and
$\bar{p}\equiv \sqrt {\beta /2m}\times p$, where
$\omega _{0}\equiv \sqrt {\mathstrut U''(q)/m}$ represents the fundamental frequency. The
potential is then assumed to be
\begin{align}
  \bar{U}(\bar{q})=\frac{1}{2!}\bar{q}^{2}+\frac{g_{3}}{3!}\bar{q}^{3},
\end{align}
where $g_{3}$ is the cubic anharmonicity of the
potential. The other variables, $V_{\mathrm{LL}}$ , $V_{\mathrm{SL}}$,
$\mu (q)$, and $\Pi (q)$, are also normalized accordingly.

\subsection{One- and two-dimensional signals}
To apply the HEOM formalism, we express the response functions in
terms of the time-propagation operator. Then, Eqs. \eqref{eq:AB} and
\eqref{eq:ABC} can be rewritten as
\begin{align}
  R(t)=\frac{i}{\hbar }\mathrm{Tr}\left\{\hat{A}\mathcal{G}(t)\hat{B}^{\times }\hat{\rho }^{\mathrm{eq}}\right\}
  \label{eq:1DIR}
\end{align}
and
\begin{align}
  R(t_{2},t_{1})=\left(\frac{i}{\hbar }\right)^{2}\mathrm{Tr}\left\{\hat{A}\mathcal{G}(t_{2})\hat{B}^{\times }\mathcal{G}(t_{1})\hat{C}^{\times }\hat{\rho }^{\mathrm{eq}}\right\}
  \label{eq:ttr2},
\end{align}
where we have employed the hyperoperator $^{\times }$ defined as
$\hat{{A}}^{\times }\hat{{B}}\equiv [\hat{{A}},\hat{{B}}]$, $\mathcal{G}(t)$ is the Green's function of
the system Hamiltonian without a laser interaction, and $\hat{\rho }_{\mathrm{eq}}$ is
the equilibrium state. The above equations represent the time
evolution of the system under laser excitation. For example,
Eq. \eqref{eq:ttr2} can be interpreted as follows. The system is
initially in the equilibrium state $\hat{\rho }^{\mathrm{eq}}$ and is then modified
as a result of the first laser pulse via the dipole
interaction by $\hat{C}$. It then propagates for time $t_{1}$ under
$\mathcal{G}(t_{1})$. The system is next excited through the second laser pulse by
$\hat{B}$ and propagates for time $t_{2}$ under $\mathcal{G}(t_{2})$. Finally, the
expectation value of the polarizability at $t_{1}+t_{2}$ is generated
through the laser pulses by $\hat{A}$.\cite{Tanimura-JPSJ-2006-75}

The classical expressions for the response functions can be obtained
from the above with the use of the Wigner
transformation.\cite{Tanimura-CP-1998-233} In this case, an arbitrary
operator $\hat{A}^{\times }(q)$ is replaced by $A'(q)(\partial /\partial p)$.
For 1D IR and 2D THz-Raman spectroscopies, we have
\begin{align}
  R_{\mathrm{IR}}^{(1)}(t)=\mu _{1}^{2}\int \mathrm{d}p\int \mathrm{d}q q\left[\mathcal{G}(t)\frac{\partial }{\partial p}W^{\mathrm{eq}}\right]
  \label{eq:1DIRW}
\end{align}
and
\begin{align}
  R_{\mathrm{RTT}}^{(3)} (t_{2},t_{1})&=\mu _{1}^{2}\Pi _{1}\int \mathrm{d}p\int \mathrm{d}q q \left[\mathcal{G}(t_{2})\frac{\partial }{\partial p}\left(\mathcal{G}(t_{1}) \left(1+\bar{\Pi }_{2}q\right)\frac{\partial }{\partial p}W^{\mathrm{eq}}(p, q)\right)\right]
  \label{eq:rtt3},\\
  R_{\mathrm{TRT}}^{(3)} (t_{2},t_{1})&=\mu _{1}^{2}\Pi _{1}\int \mathrm{d}p\int \mathrm{d}q q \left[\mathcal{G}(t_{2})\left(1+\bar{\Pi }_{2}q\right)\frac{\partial }{\partial p}\left(\mathcal{G}(t_{1})\frac{\partial }{\partial p}W^{\mathrm{eq}}(p, q)\right)\right],
  \label{eq:trt3}
\end{align}
and
\begin{align}
  R_{\mathrm{TTR}}^{(3)} (t_{2},t_{1})&=\mu_{1}^{2}\Pi _{1}\int \mathrm{d}p\int \mathrm{d}q \left(q+\frac{1}{2}\bar{\Pi }_{2}q^{2}\right) \left[\mathcal{G}(t_{2})\frac{\partial }{\partial p}\left(\mathcal{G}(t_{1})\frac{\partial }{\partial p}W^{\mathrm{eq}}(p, q)\right)\right]
  \label{eq:ttr3},
\end{align}
respectively. Here, $\bar{\Pi }_{2}\equiv \Pi _{2}/\Pi _{1}$ is the relative
intensity of the nonlinear polarizability. The Green's function $\mathcal{G}(t)$
is now expressed in terms of the classical Liouvillian as
$\mathcal{G}(t)=\exp \left[-\mathcal{\hat{L}}_{\mathrm{cl}}t\right]$, and $W^{\mathrm{eq}}(p, q)$ is the equilibrium
distribution. In order to apply the HEOM formalism, we express the
time-dependent Wigner function $W(p,q;t)$, such as
$W(p,q;t_{1})=\mathcal{G}(t_{1}){\partial}W^{eq}(p,q)/ {\partial p}$ and
$W(p,q;t_{1}+t_{2})=\mathcal{G}(t_{2}){\partial}W(p,q; t_{1})/{\partial p}$, in terms of the HEOM
member $W^{(n)}(p,q;t_{1})$ and $W^{(n)}(p,q;t_{1}+t_{2})$, and the determine its
time evolution through Eqs.\eqref{eq:heom1} and
\eqref{eq:heom2}.\cite{Steffen-JPSJ-2000-69,Tanimura-JPSJ-2000-69,
  Kato-JCP-2002-117,Kato-JCP-2004-120,Tanimura-JPSJ-2006-75} In the
HEOM formalism, the equilibrium distribution, $W^{\mathrm{eq}}(p,q)$, is also
expressed using the HEOM elements evaluated from the steady-state
solution of Eqs.  \eqref{eq:heom1} and \eqref{eq:heom2}. In the strong
coupling case, we employ an eigenfunctional representation of the
momentum space for numerical convenience, as discussed in Appendix
\ref{sec:hermite_rep}, while, in other cases, we solve
Eqs. \eqref{eq:heom1} and \eqref{eq:heom2}.  Note that, as shown in
Ref. \onlinecite{Tanimura-PRA-1991-43}, the hierarchal Fokker-Planck
approach is equivalent to the generalized Langevin approach.  However,
because the Fokker-Planck approach does not require sampling of system
trajectories, unlike the Langevin approach, it is numerically
advantageous, especially for calculating nonlinear response functions,
for which the trajectories are unstable.

\section{Results and Discussion}
\label{sec:results_and_discussion}
Because the 2D profiles of the signal must be constructed from complex
motion in a complicated manner, the analysis of the signal profile is
not straightforward.  Nevertheless, properly accounting for the
components of signals to these profiles allows us to perform a
detailed analysis of the inter-molecular vibrational motion on the
basis of both experiential and theoretical results. Model-based
studies of the 2D profiles of signals are helpful to identify the
underlying physical mechanisms, because it is necessary to capture the
essential features of the inter-molecular motion in order to reproduce
the complex 2D profile from a simple model. Analyses of this kind have
employed LL + SL BO models for the 2D IR\cite{Ishizaki-JCP-2006-125,
  Ishizaki-JPC-2007-111, Tanimura-ACR-2009-42, Sakurai-JPCA-2011-115}
and 2D Raman cases.\cite{Steffen-JPSJ-2000-69,Tanimura-JPSJ-2000-69,
  Kato-JCP-2002-117,Kato-JCP-2004-120,Tanimura-JPSJ-2006-75} However,
their applicability has not been fully explored because of the limited
availability of experimental and numerical data.  2D THz-Raman
measurements, which are applicable not only to the study of liquids,
but also to the problem of distinguishing optical processes through
use of RTT, TRT, and TTR measurements, provide the opportunity to
explore the possibilities of model-based analysis of 2D profiles.
Here, we examine the LL+SL BO model and use it to reproduce all three
THz-Raman signal profiles obtained from full MD simulations. This is
done by choosing the parameter values of the model so as to realize
the best agreement between the signal profiles provided by the model
and the MD simulation.  Before fitting the model to the MD signals, we
demonstrate a general aspect of 2D THz-Raman signals using the SL BO
model and the optical Liouville paths in a simple case. This serves as
a guid to subsequent analysis.

\subsection{General aspects of 2D THz-Raman signals}
\newcommand{\CorAH}[0]{\bar{R}_{\mathrm{AH}}} \newcommand{\CorRTT}[0]{\bar{R}_{\mathrm{RTT}}}
\newcommand{\CorTRT}[0]{\bar{R}_{\mathrm{TRT}}} \newcommand{\CorTTR}[0]{\bar{R}_{\mathrm{TTR}}}
\newcommand{\CorAHp}[0]{\CorAH(t_{2}, t_{1})}
\newcommand{\CorRTTp}[0]{\CorRTT(t_{2}, t_{1})}
\newcommand{\CorTRTp}[0]{\CorTRT(t_{2}, t_{1})}
\newcommand{\CorTTRp}[0]{\CorTTR(t_{2}, t_{1})}
Here, we elucidate several aspects of the signal components in 2D THz-Raman
spectroscopy using the SL BO model and optical Liouville paths.  As
explained in Sec. \ref{sec:introduction}, we can express the dipole
and polarizability in terms of the collective coordinate $\hat{\bm{q}}$ as
$\hat{\bm{\mu }}(\hat{\bm{q}})=\mu _{1}\hat{\bm{q}} $ and $\hat{\bm{\Pi }}(\hat{\bm{q}})=\Pi _{1}\hat{\bm{q}}+\Pi _{2}\hat{\bm{q}}^{2}/2$, respectively.
If the inter-molecular modes are both Raman and THz active, the three
THz-Raman signals are expressed as
\begin{align}
  R_{\mathrm{RTT}}^{(3)}(t_{2},t_{1})&=\mu _{1}^{2}\Pi _{1}\left[\CorAHp+\frac{\bar{\Pi }_{2}}{2}\CorRTTp\right]\label{eq:rtt},\\
  R_{\mathrm{TRT}}^{(3)}(t_{2},t_{1})&=\mu _{1}^{2}\Pi _{1}\left[\CorAHp+\frac{\bar{\Pi }_{2}}{2}\CorTRTp\right]\label{eq:trt},
\end{align}
and
\begin{align}
  R_{\mathrm{TTR}}^{(3)}(t_{2},t_{1})&=\mu _{1}^{2}\Pi _{1}\left[\CorAHp+\frac{\bar{\Pi }_{2}}{2}\CorTTRp\right]\label{eq:ttr},
\end{align}
with $\bar{\Pi }_{2}\equiv \Pi _{2}/\Pi _{1}$, where the anharmonic component is expressed as
\begin{align}
  \CorAHp&\equiv -\frac{1}{\hbar ^{2}}\left\langle [[\hat{\bm{q}}(t_{12}),\hat{\bm{q}}(t_{1})],\hat{\bm{q}}(0)]\right\rangle. \label{eq:ah}
\end{align}
Similarly the nonlinear polarizability components are given by
\begin{align}
  \CorRTTp&\equiv -\frac{1}{\hbar ^{2}}\left\langle [[\hat{\bm{q}}(t_{12}),\hat{\bm{q}}(t_{1})],\hat{\bm{q}}^{2}(0)]\right\rangle \label{eq:cor_rtt},\\ 
  \CorTRTp&\equiv -\frac{1}{\hbar ^{2}}\left\langle [[\hat{\bm{q}}(t_{12}),\hat{\bm{q}}^{2}(t_{1})],\hat{\bm{q}}(0)]\right\rangle \label{eq:cor_trt},
\end{align}
and
\begin{align}
  \CorTTRp\equiv -\frac{1}{\hbar ^{2}}\left\langle [[\hat{\bm{q}}^{2}(t_{12}),\hat{\bm{q}}(t_{1})],\hat{\bm{q}}(0)]\right\rangle\label{eq:cor_ttr}.
\end{align}

In the harmonic LL BO case, the above THz-Raman signals can be
calculated analytically, and we have $\CorAH=0$, $\CorRTT=0$,
\begin{align}
  \CorTRTp=-\frac{1}{\hbar^2}C(t_{1})C(t_{2}),
  \label{eq:cor_trtHBO}
\end{align}
and
\begin{align}
  \CorTTRp = -\frac{1}{\hbar^2} C(t_{1}+t_{2})C(t_{2}),
  \label{eq:cor_ttrHBO}
\end{align}
where $C(t)\equiv \langle [q(t), q]\rangle $ is the first-order response function of the
harmonic BO system. For an isolated oscillator with frequency
$\omega$, we have $C(t)=\hbar \sin (\omega t)/2 m \omega$.  As explained in
Sec. \ref{sec:introduction}, the term $\CorAH$ vanishes in the
harmonic case because there is an odd number of Gaussian integrals
involved in the response function.\cite{Okumura-JCP-1996-105,
  Okumura-CPL-1997-277} Moreover, the term $\CorRTT$ vanishes because
of the cancelation of possible optical Liouville paths, as explained
in Appendix \ref{sec:rtt}. While $\CorAH$ becomes large for large
anharmonicity $g_3$,\cite{Okumura-JCP-1996-105, Okumura-CPL-1997-277}
the contribution of $\CorRTT$ remains small due to this cancelation.
Thus, in the anharmonic case, we can estimate $\CorAH$ from the RTT
measurement. Then, by subtracting $\CorAH$ from the TRT and TTR
signals, $R_{\mathrm{TRT}}^{(3)}$ and $R_{\mathrm{TTR}}^{(3)}$, we can evaluate $\CorTRT$ and
$\CorTTR$ separately. Because each contribution arises from the
corresponding optical process, we can elucidate the key features of
the inter-molecular motion from them. In the 2D Raman case,
contrastingly, because all of the contributions appear together in a
single observable, we cannot carry out such analysis.

To more clearly elucidate the characteristic of the 2D THz-Raman
signals, in the figures, we display the 2D profiles of the $\CorTRT$,
$\CorTTR$, and $\CorAH$ components separately in the slow modulation
case ($\zeta = 1.0\ \omega _{0}$ and $\gamma = 0.5\ \omega _{0}$) and the fast
modulation case ($\zeta = 0.49\ \omega _{0}$ and $\gamma =\infty $), as obtained
from the SL BO model ($V_{\mathrm{LL}}=0$, $V_{\mathrm{SL}}=1$) with $\omega _{0}=600$ $\mathrm{cm}^{-1}$ and
$T=300\ \mathrm{K}$. Note that, because the effective system-bath coupling
strength becomes weaker as the modulation becomes faster, we change
both $\gamma $ and $\zeta $ to elucidate a pure non-Markovian
effect.\cite{Tanimura-JCP-accepted} Although the contribution is
minor, we plot the $\CorRTT$ component in Appendix \ref{sec:rtt}. The
relative intensities of each component evaluated in the harmonic and
anharmonic SL BO cases are presented in Table. \ref{tab:demo}. We then
analyze each profile using the optical Liouville paths (the
double-sided Feynman diagrams).
\begin{table*}[!tb]
  \caption{\label{tab:demo} Relative intensities of the signal
    components for the SL BO model in the harmonic ($g_3=0$) and
    anharmonic ($g_3=0.3$) cases. The intensities are estimated from
    the maximum peak value of the signal normalized with respect to
    the values of $\CorTRT$. Although the SL interaction gives rise
    to the contribution of $\CorRTT$ to the RTT signal, its intensity is
    weaker than that of $\CorAH$ if we include the prefactor
    $\bar \Pi_2$. This situation does not change even if we increase the
    strength of the anharmonicity.\\ }
  \begin{tabular}{cc|@{\hspace{4mm}}cccc}
    \hline \hline
    Potential & Modulation & $\CorAH$ & $\CorRTT$$^{a)}$ & $\CorTRT$ & $\CorTTR$ \\
    \hline
    harmonic   & slow &  ---   & $0.18$ &   $1$  & $0.99$ \\
    harmonic   & fast &  ---   & $0.11$ & $0.85$ & $0.81$ \\
    anharmonic & slow & $0.12$ & $0.18$ & $1.01$ & $1.00$ \\
    anharmonic & fast & $0.14$ & $0.13$ & $0.85$ & $0.81$ \\
    \hline \hline
  \end{tabular}
  \begin{flushleft}
    $^{a)}$ The 2D profile of the RTT component is presented in Appendix \ref{sec:rtt}. \\
  \end{flushleft}
\end{table*}

\begin{figure}[bthp]
  \centering
  \includegraphics[scale=1.0]{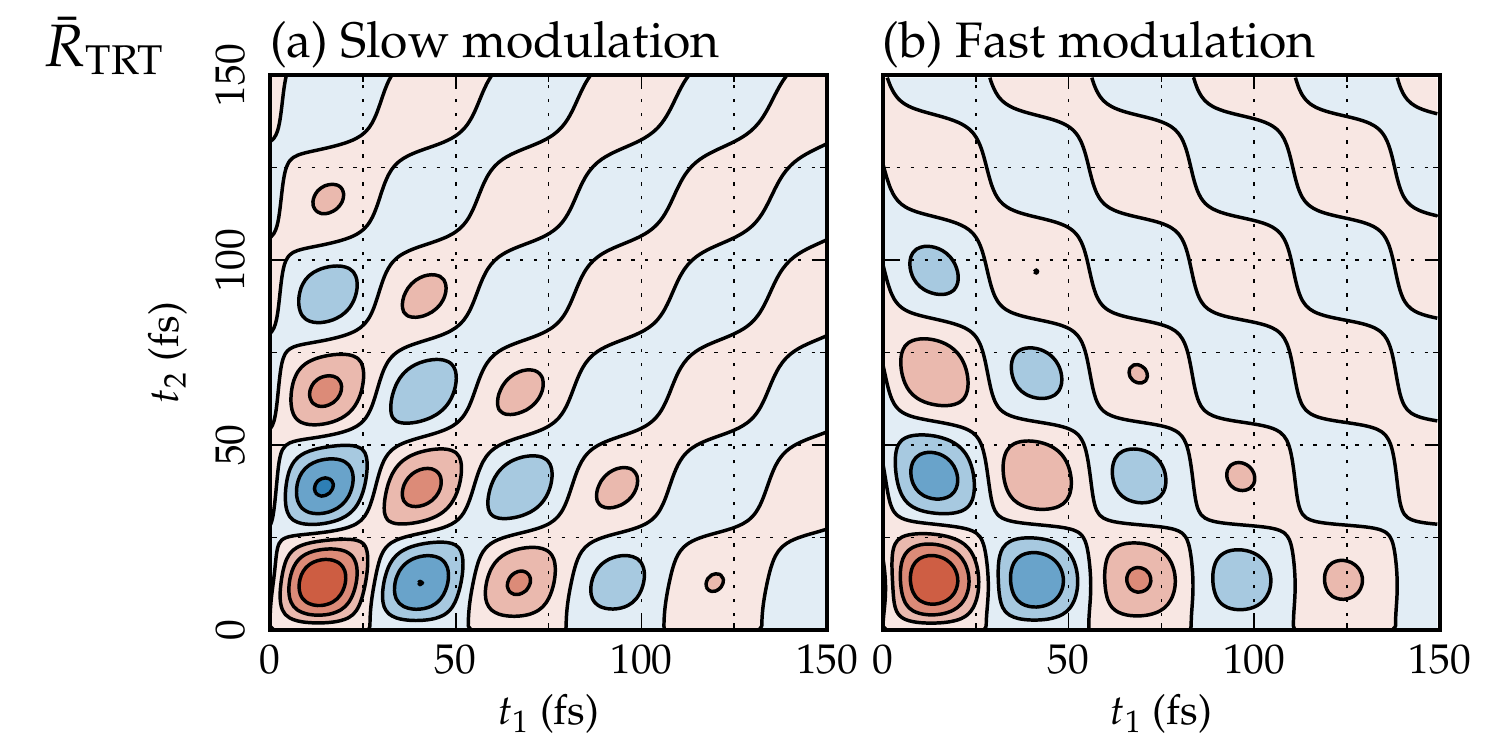}
  \caption{The 2D profiles of the $\CorTRT$ component calculated using
    the harmonic SL BO model ($g_{3}=0$) for the (a) slow and (b) fast
    modulation cases. Contours in red and blue represent positive
    and negative values, respectively. }
  \label{fig:h_trt}
\end{figure}
\begin{figure}[bthp]
  \centering
  \includegraphics[scale=1.7]{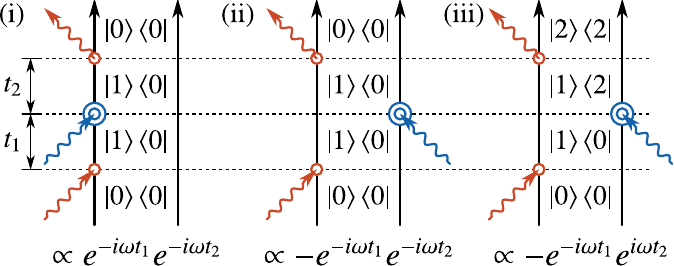}
  \caption{The Liouville paths involved in $\CorTRT$. The red circles and
    the blue double circles represent single and double quantum
    transitions, respectively. The three other paths, which are the
    Hermitian conjugates of the above-mentioned  paths, are not presented. These
    conjugate paths can be obtained by exchanging the left and right
    arrows.}
  \label{fig:diagrams_121}
\end{figure}
\subsubsection{The TRT component, $\CorTRTp$}
In Fig. \ref{fig:h_trt}, we present the $\CorTRT$ component for (a)
the slow and (b) the fast modulation cases calculated using the
harmonic SL BO model ($g_{3}=0$).  It is seen that the peak profiles are
symmetric along the $t_{1}=t_{2}$ line, as can be deduced from the
analytical expression for the LL case, Eq. \eqref{eq:cor_trtHBO}. The
peaks in the slow modulation case stretch in the $t_{1}-t_{2}=2n\pi /\omega $
direction, whereas those in the fast modulation case stretch in the
$t_{1}+t_{2}=2n\pi /\omega $ direction, where $\omega $ is the fundamental frequency and
$n$ an integer. In the slow modulation case, there are elongated peaks
called ''echo peaks'' along the $t_{1}=t_{2}$ direction.

Although our simulation results are fully classical, these profiles
can be interpreted easily using the quantum Liouville paths for the
TRT process depicted in Fig. \ref{fig:diagrams_121}. There, an energy
eigenstate of the harmonic potential is denoted by $|n\rangle $, and we have
depicted cases starting from the vibrational ground state as examples.
The dipole operator $\mu _{1}\hat{q}$, represented by the red circles in the
diagram, converts the state $|n\rangle $ into $|n+1\rangle $ and $|n-1\rangle $ through
single quantum (SQ) excitations, while the nonlinear polarizability
operator $\Pi _{2}\hat{q}^{2}$, represented by the blue double circles, converts the
state $|n\rangle $ into $|n+2\rangle $ and $|n-2\rangle $ through double quantum (DQ)
excitations or maintains the same state $|n\rangle $ through the zero quantum
(ZQ) excitation. Because the final state, appearing after the last
laser interaction, must be a population state $|n\rangle \langle n|$, due to the
trace operation involved in the response function, the possible
processes are limited to the cases depicted in
Figs. \ref{fig:diagrams_121}(i)-(iii) and their conjugate
diagrams. While the double circle in the diagram (i) involves the
transition $(a^{\dagger }a + aa^{\dagger })|n+1\rangle =(2n+3)|n+1\rangle $,
that in the diagram (ii) involves the transition
$(a^{\dagger }a + aa^{\dagger })|n\rangle =(2n+1)|n\rangle $. Thus, although
these paths have same phases with opposite signs, they do not cancel.
These six components constitute the signal expressed in
Eq. \eqref{eq:cor_trtHBO} in the isolated oscillator case.

The phases for the paths (i) and (ii) in
Fig. \ref{fig:diagrams_121}(i), expressed as
$\exp[-i\omega (t_{1}+t_{2})]$, are the same because the net transition for
these paths is a ZQ transition, while that for path (iii), expressed
as $\exp[-i\omega (t_{1}-t_{2})]$, is different because the net transition
for this path is a DQ transition.  In the slow modulation case, the
signal in Fig. \ref{fig:diagrams_121} (iii) can rephase and become
strong along the $t_{1}=t_{2}$ direction, but in the fast modulation case,
coherence is lost due to rapid changes in the fundamental frequency,
as illustrated in Fig. \ref{fig:homo_inhomo}, and the signal decays
quickly.  Thus, we observe the echo signal in the TRT component in the
slow modulation case, whereas we observe the chain of peaks along the
$t_{1}+t_{2}=2n\pi/\omega$ direction in the fast modulation case, due to
the processes depicted in Figs. \ref{fig:diagrams_121} (i) and (ii).

\begin{figure}[bthp]
  \centering
  \includegraphics[scale=1.0]{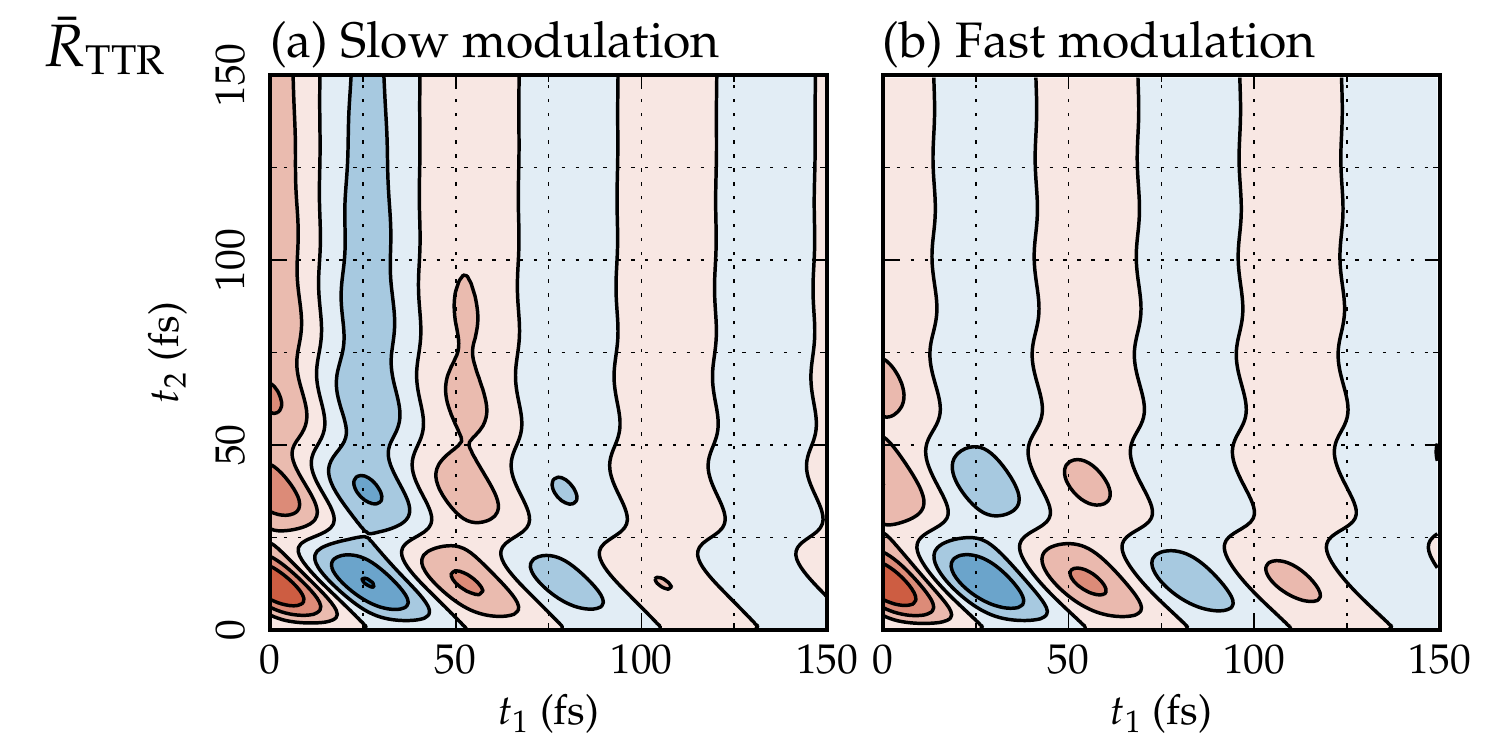}
  \caption{The 2D profiles of the $\CorTTR$ component calculated using
    the harmonic SL BO model ($g_{3}=0$) for the (a) slow and (b) fast
    modulation cases. Contours in red and blue represent positive
    and negative values, respectively. }
  \label{fig:h_ttr}
\end{figure}
\begin{figure}[bthp]
  \centering
  \includegraphics[scale=1.7]{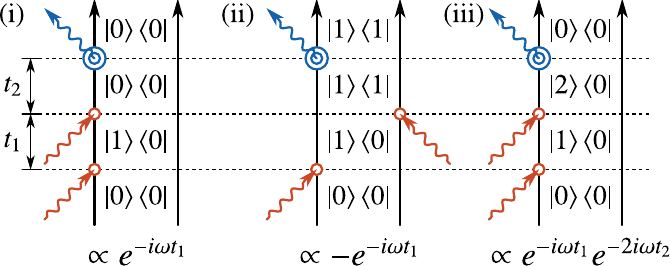}
  \caption{The Liouville paths involved in $\CorTTR$. The red circles and
    the blue double circles represent single and double quantum
    transitions, respectively. The three other paths, which are the
    Hermitian conjugates of the above paths, are not presented. }
  \label{fig:diagrams_112}
\end{figure}

\subsubsection{The TTR component, $\CorTTRp$}
We now discuss the $\CorTTR$ component, presented in
Fig. \ref{fig:h_ttr}. The prominent features of this signal are the
appearance of elongated peaks parallel to the $t_{2}$ axis and the
appearance of peaks along the $t_{1}+2t_{2}=2n\pi /\omega $ direction for small
values of $t_{2}$. The optical Liouville paths for $\CorTTR$ are
presented in Fig. \ref{fig:diagrams_112}. Three other paths can be
obtained by exchanging the left and right arrows. These six components
constitute the signal expressed in Eq. \eqref{eq:cor_ttrHBO} in the
isolated oscillator case.

We observe the population states $|n\rangle \langle n|$ during the $t_{2}$ period of
Figs. \ref{fig:diagrams_112}(i) and \ref{fig:diagrams_112} (ii) and
the coherent states $|n+2\rangle \langle n|$ created by the two SQ excitations
during the $t_{2}$ period of Figs. \ref{fig:diagrams_112}(iii).  While
all of the diagrams in Fig. \ref{fig:diagrams_112} exhibit the
oscillation $\exp [-i\omega t_{1}]$ during the $t_{1}$ period, only that in
Fig. \ref{fig:diagrams_112}(iii) exhibits the oscillation
$\exp [-2i \omega t_{2}]$ during the $t_{2}$ period. Due to the SL interaction, the
high-frequency oscillation of the coherent state appearing in
Fig. \ref{fig:diagrams_112}(iii) decays quickly, while the population
state appearing in Fig. \ref{fig:diagrams_112}(ii) remains for a long
time.  Thus, we observe the peaks along the $t_{1}+2t_{2}=2n\pi /\omega $ direction
for a short $t_{2}$ period, whereas elongated peaks appear along the $t_{2}$
direction.

\begin{figure}[bthp]
  \centering
  \includegraphics[scale=1.0]{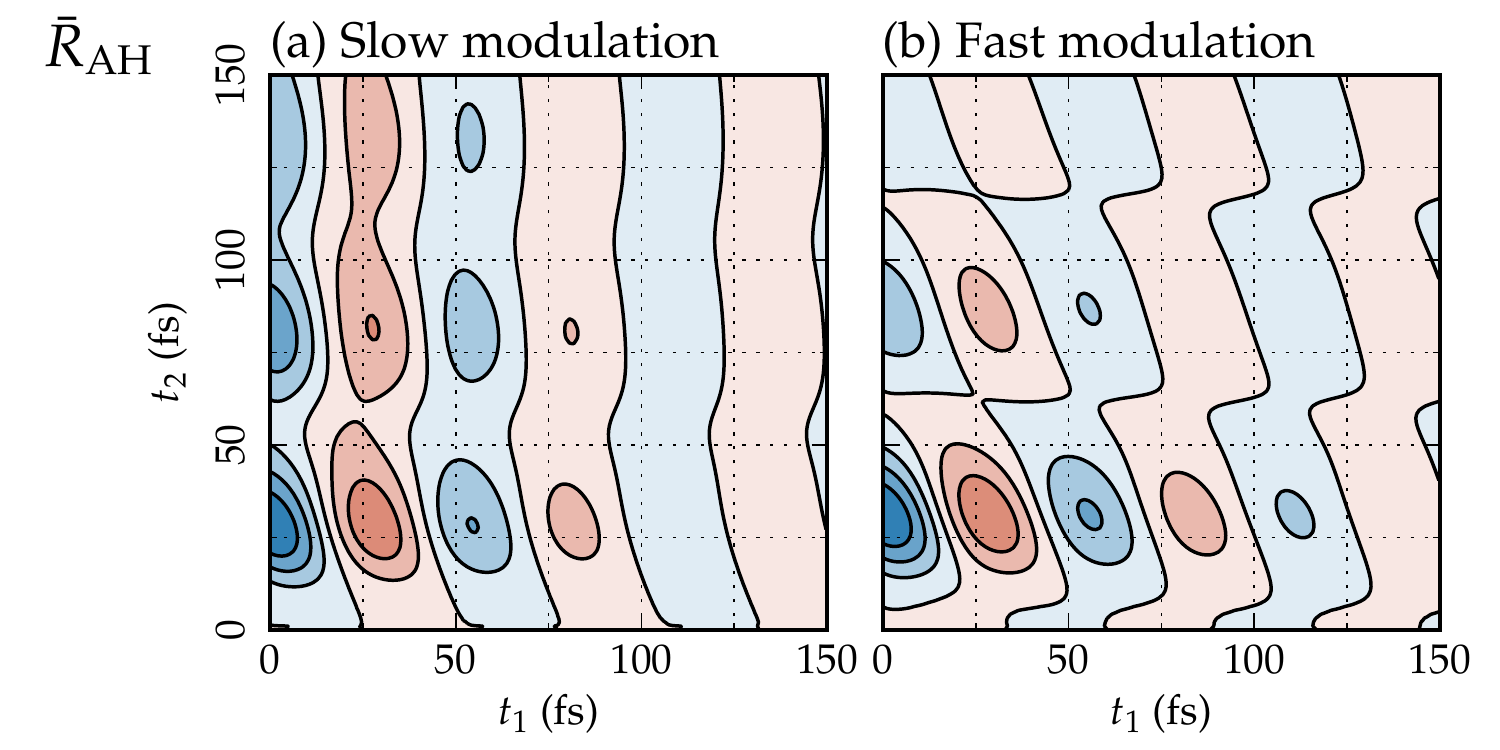}
  \caption{The 2D profiles of the $\CorAH$ component calculated using
    the anharmonic SL BO model ($g_{3}=0.3$) for the (a) slow and (b)
    fast modulation cases. Contours in red and blue represent
    positive and negative values, respectively. }
  \label{fig:ah_ah}
\end{figure}

\subsubsection{The AH component, $\CorAHp$}
The contribution of the component $\CorAH$ arises only in the
anharmonic case. In Fig. \ref{fig:ah_ah}, we display the signal for
$g_{3}=0.3$ in the (a) slow and (b) fast modulation cases,
respectively. Although the definition of the response function is
different, the optical Liouville paths for $\CorAH$ are similar to
those for $\CorTRT$ and $\CorTTR$. To explain the reason for
this, we consider energy eigenstates in the case of an anharmonic
potential $|n'\rangle $ with eigenenergy $\omega _{n'}$. The fundamental frequency
between $|0'\rangle $ and $|1'\rangle $ is denoted by $\omega '=\omega _{1'}-\omega _{0'}$, whereas
that between $|1'\rangle $ and $|2'\rangle $ is denoted by $\omega '-\delta =\omega _{2'}-\omega _{1'}$.  In
addition to such frequency shifts, the anharmonicity changes the roles
of the dipole and polarizability operators. While the $\CorAH$
component involves only $\mu _{1}\hat{q}$ and $\Pi _{1}\hat{q}$, they can induce ZQ and
DQ transitions, in addition to the SQ transition, because
$\langle n' |\hat{q}|n'\rangle $, $\langle n'|\hat{q}|n'+2'\rangle $, and
$\langle n'+2'|\hat{q}|n'\rangle $ are all nonzero in the anharmonic case. Because
both THz and Raman laser pulses can induce ZQ and DQ transitions, the
diagrams for $\CorAH$ include all of the diagrams in
Figs. \ref{fig:diagrams_121} and \ref{fig:diagrams_112}, while the
resonant frequency for the $t_{1}$ period becomes $\omega '$, and that
for the $t_{2}$ period is $0$ , $\omega '$, $\omega '-\delta $, or
$2\omega '-\delta $.  The rephasing processes depicted in
Fig. \ref{fig:diagrams_121} (iii) occur only rarely, even in the slow
modulation case, due to the anharmonicity, and this contribution can
therefore be ignored. For this reason, the 2D profiles of the $\CorAH$
component are similar to those of $\CorTTR$ and $\CorTTR$, but without
the echo peaks described by Fig. \ref{fig:diagrams_121} (iii). As in
the TTR case, the contribution from the diagram depicted in
Fig. \ref{fig:diagrams_112} (iii) decays quickly. Thus, we observe
elongated peaks in the $t_{2}$ direction that arise from population decay
during the $t_{2}$ period.

\subsection{The MD and fitted results}
In order to concretely study hydrogen-bonding dynamics in 2D THz-Raman
spectroscopy, we selected formamide, water, and methanol. While water
and methanol exhibit high-frequency librational motion arising from
hydrogen bonding, formamide exhibits only low-frequency
inter-molecular motion. For each liquid, we explore the parameter
values of the LL+SL BO model to fit the RTT, TRT and TTR signals to
the MD simulation results.  We fixed the anharmonicity to $g_{3}=0.1$,
while we varied the nonlinear polarizability $\bar{\Pi }_{2}$. This was done
because the intensity of the $\CorAH$ component is proportional to
$g_{3}$,\cite{Okumura-JCP-1996-105, Okumura-CPL-1997-277} and only the
ratio of $g_{3}$ and $\bar{\Pi }_{2}$ is important to elucidate the roles of
anharmonicity and non-linear polarizability. Because the coupling
strengths of the LL and SL interactions are determined by $\zeta V_{\mathrm{LL}}$ and
$\zeta V_{\mathrm{SL}}$, respectively, we also fixed $V_{\mathrm{SL}}=1$ and varied $\zeta $ and
$V_{\mathrm{LL}}/V_{\mathrm{SL}}$ in the fit.  The best fits of these parameters are presented
in Table \ref{tab:params}. Note that because the SL interaction
increases the effective system frequency,
\cite{Steffen-JPSJ-2000-69,Tanimura-JPSJ-2000-69} the fitted $\omega _{0}$ do
not correspond to the resonant frequency estimated from the 1D IR
results obtained from the MD simulation.
\begin{table*}[!tb]  
  \caption{\label{tab:params}The fitted parameter values of the LL+SL
    BO model for formamide, water and methanol liquid. Here, have
    fixed $g_{3}=0.1$, $\mu _{1}=1$, $\Pi _{1}=1$, and $V_{\mathrm{SL}}=1$.\\}
\begin{tabular}{c|ccccccc}
  \hline \hline
  Molecule & $\omega _{0}\ (\text{cm}^{-1})$ & $\bar{\Pi }_{2}$ & $\zeta /\omega _{0}$ & $\gamma /\omega _{0}$  & $V_{\mathrm{LL}}/V_{\mathrm{SL}}$ \\
  \hline
  Formamide & $60$              & $0.04$ & $20$  & $\infty $   & $0$  \\    
  Water     & $450$             & $0.07$ & $6.0$ & $0.7$ & $0.01$ \\
  Methanol  & $540$             & $0.07$ & $2.1$ & $0.4$ & $0.01$ \\
  \hline \hline
\end{tabular}
\end{table*}

\subsubsection{Formamide}
\begin{figure}[p]
  \centering
  \includegraphics[scale=1.0]{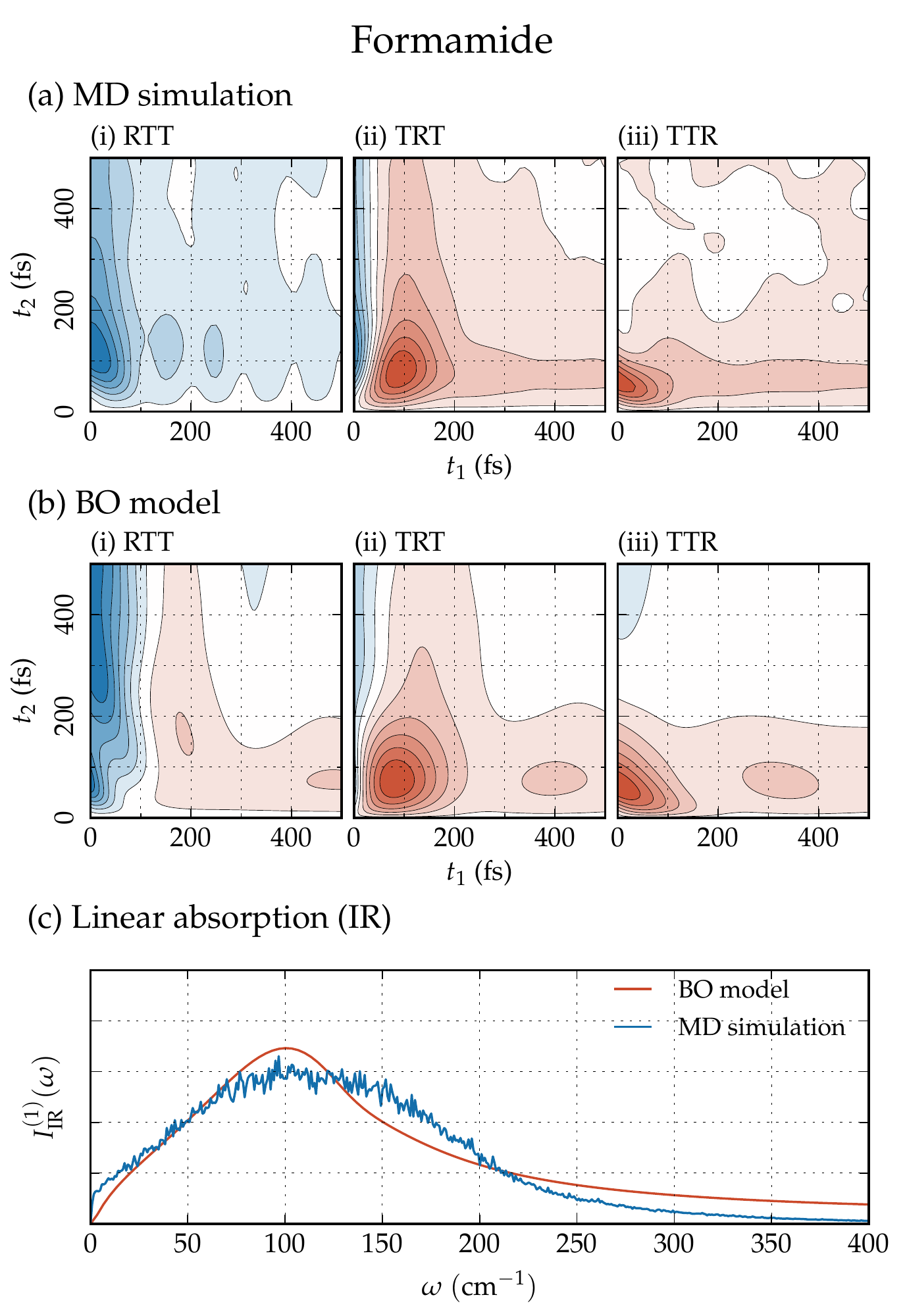}
  \caption{The 2D signals for formamide obtained from (a) the MD
    simulation and (b) the LL+SL BO model for the (i) RTT, (ii) TRT,
    and (iii) TTR measurements, respectively. The fitted parameter
    values of the LL+SL BO model are presented in Table
    \ref{tab:params}. Using these values, we also calculated the 1D IR
    signal, which is presented with the MD results in (c).}
  \label{fig:formamide}
\end{figure}

In Fig. \ref{fig:formamide}, we display (a) the MD results and (b) the
fitted results from the LL+SL BO model for the (i) RTT, (ii) TRT, and
(iii) TTR cases of liquid formamide. It is seen that the LL+SL BO
results capture the essential features of the 2D THz-Raman
spectra. The characteristic feature of the MD signals for formamide is
the elongation of the peaks along the $t_{2}$ axis observed in
Figs. \ref{fig:formamide}(a-i) and (a-ii). By adapting the analysis
applied to $\CorAH$ and $\CorTRT$, we conclude that this elongation
arises from the slow population decay during the $t_{2}$ period described
by the diagrams in Figs. \ref{fig:diagrams_121} (i),
\ref{fig:diagrams_121} (ii), \ref{fig:diagrams_112} (i), and
\ref{fig:diagrams_112} (ii). The echo peaks do not appear in
Figs. \ref{fig:formamide}(a-ii) and (b-ii), because the modulation in
this case is so fast that dephasing cannot take place. In the TTR case
depicted in Figs. \ref{fig:formamide}(a-iii) and (b-iii), the peaks
along the $t_{1}+2t_{2}=2n\pi /\omega $ direction appear for small values of $t_{1}$ and
$t_{2}$ due to the contribution of the diagram presented in
Fig. \ref{fig:diagrams_112} (iii). The similarity between the MD and
model results indicates that the collective modes are subject to fast
frequency modulation, as illustrated in Fig. \ref{fig:homo_inhomo}(a).

\subsubsection{Water and Methanol}
\begin{figure}[p]
  \centering
  \includegraphics[scale=1.0]{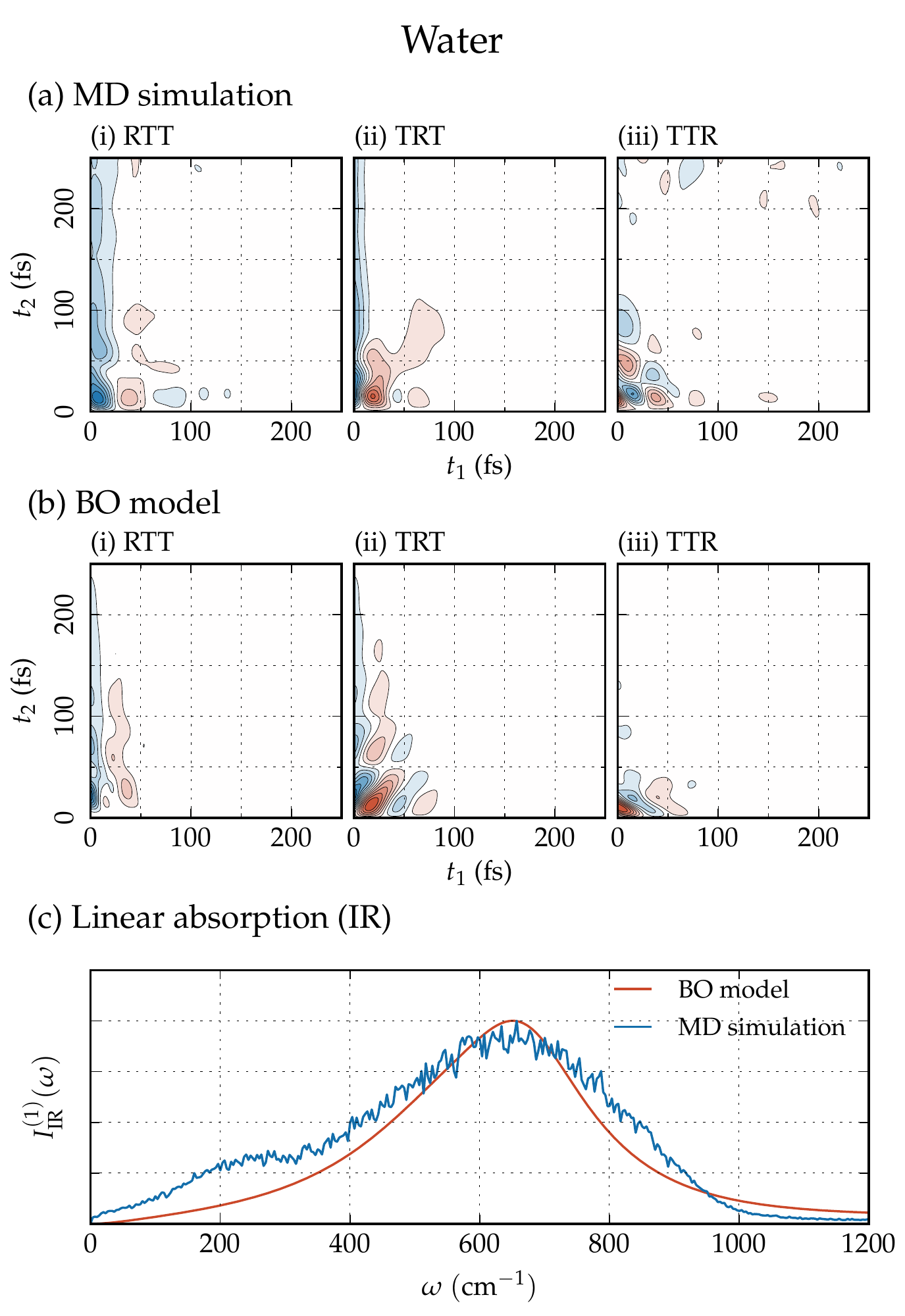}
  \caption{ The 2D signal for water obtained from (a) the MD
    simulation and (b) the LL+SL BO model for the (i) RTT, (ii) TRT,
    and (iii) TTR measurements, respectively. The fitted parameter
    values of the LL+SL BO model are presented in Table
    \ref{tab:params}. Using these values, we also calculated the 1D IR
    signal, which is presented with the MD results in (c).}
  \label{fig:water}
\end{figure}
\begin{figure}[p]
  \centering
  \includegraphics[scale=1.0]{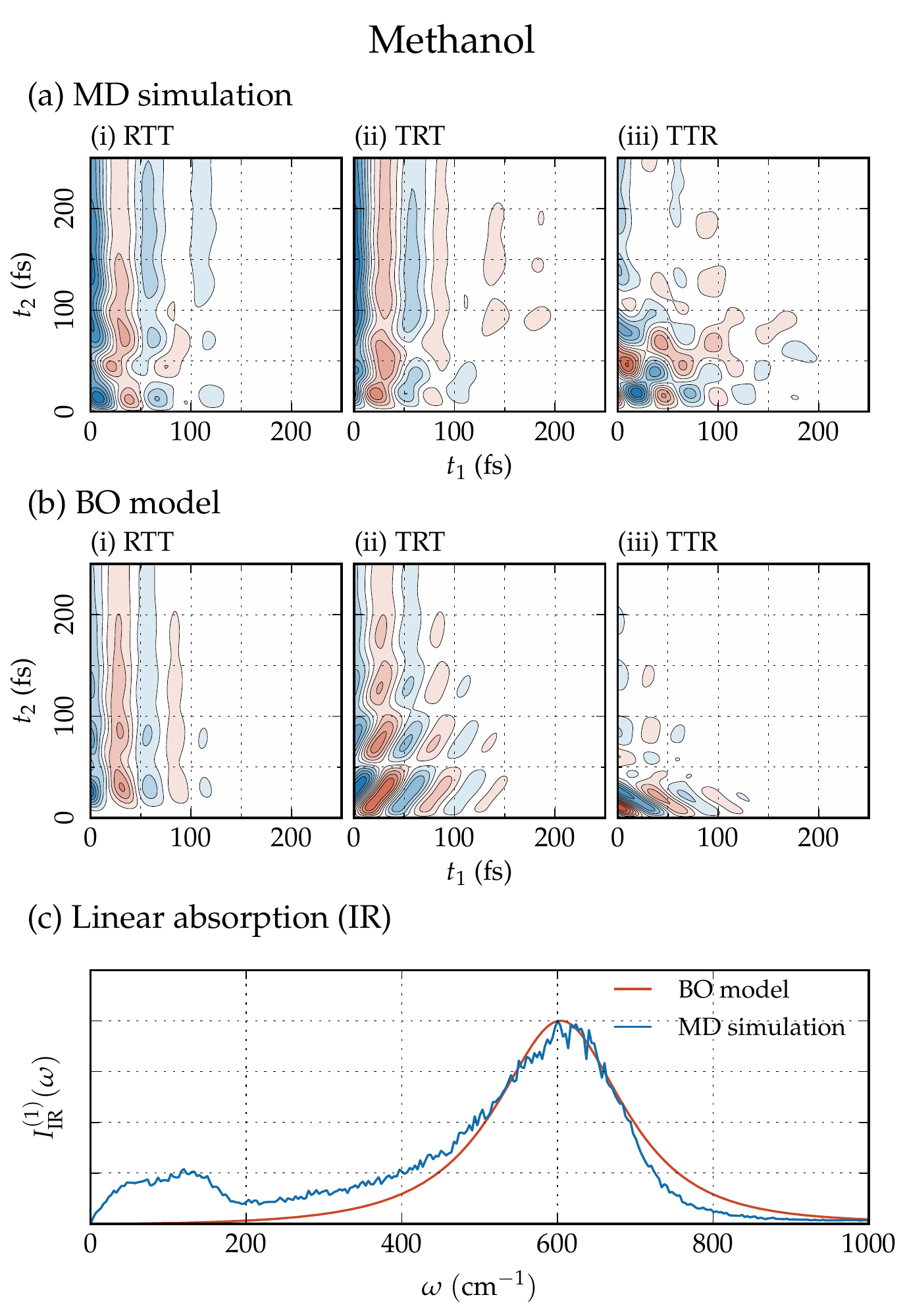}
  \caption{ The 2D signal for methanol obtained from (a) the MD
    simulation and (b) the LL+SL BO model for the (i) RTT, (ii) TRT,
    and (iii) TTR measurements, respectively. The fitted parameter
    values of the LL+SL BO model are presented in Table
    \ref{tab:params}. Using these values, we also calculated the 1D IR
    signal, which is presented with the MD results in (c).}
  \label{fig:methanol}
\end{figure}

In Figs. \ref{fig:water} and \ref{fig:methanol}, we display (a) the MD
results and (b) the LL+SL BO results for the (i) RTT, (ii)
TRT, and (iii) TTR cases of liquid water and methanol,
respectively.  We reproduced all of the 2D profiles obtained from the
MD simulation for these liquids using the LL+SL BO model with the
parameter values listed in Table \ref{tab:params}. In these liquids,
the vibrational modes near $\omega =600 \text{ cm}^{-1}$ are the librational
modes.\cite{Bosma-JCP-1993-98, Larsen-JCP-2006-125}

The characteristic features of the MD signals for these two liquids
are the elongation of the peaks along the $t_{2}$ axis in the RTT and TRT
signals and the echo peaks along the $t_{1}=t_{2}$ line in the TRT
signals. The existence of these peaks indicates that the collective
modes are inhomogeneously distributed, as depicted in
Fig. \ref{fig:homo_inhomo} (b), due to slow frequency
modulation. Because water and methanol exhibit high-frequency
librational motion caused by hydrogen bonds, we observe an oscillatory
feature in the 2D signal. This contrasts with the formamide signal,
which decays quickly.

\section{Conclusion}
\label{sec:conclusion}
Using a Brownian oscillator (BO) model with a linear-linear (LL) and
square-linear (SL) system-bath interaction, we analyzed the 2D RTT,
TRT, and TTR signals for formamide, water, and methanol liquids
obtained from full MD simulations.  The classical hierarchal equations
of motion approach was used to calculate the 2D signals with the LL+SL
BO model under non-perturbative and non-Markovian conditions. By
fitting the anharmonicity of the potential, the LL and SL coupling
strength, and the inverse noise correlation time of the LL+SL BO model
to the results of the MD simulations, we reproduced each of the
simulated signal profiles by capturing their characteristic features.
We found that the profile of formamide liquid can be accounted for by
homogeneously distributed oscillators, whereas the profiles of water
and methanol liquids can be accounted for by inhomogeneously
distributed oscillators.  Due to hydrogen bond interactions, water and
methanol exhibit oscillating echo peaks.  We were able to describe the
two cases as the cases of fast and slow modulation of the anharmonic
LL+SL BO model.  The key feature of the present model that allows it
to account for the simulated signal is the existence of the SL
interaction. We were able to describe the complex inter-molecular
motion with this simple model because it is capable of describing the
collective modes under conditions ranging from homogeneous to
inhomogeneous in a unified manner through variation of the noise
correlation time. The success of the present study indicates that the
LL+SL BO model captures the essence of the inter-molecular motion.

Finally, we briefly discuss some extensions of the present
study. First, note that there does exist some discrepancy between the
MD simulation results and the LL+SL BO model results. It should not be
too difficult to decrease this discrepancy. For example, in the case
of water, we should be able to improve the description of the signal
profiles by introducing the second mode, which is $220$ $\mathrm{cm}^{-1}$, in the
BO model. The significance of the fitted frequencies, $\omega _{0}$, should be
clarified through the normal mode analysis of MD
simulations. Secondly, as shown in previous studies, the LL+SL BO
model can be applied not only to 2D THz-Raman spectroscopy, but also
to 2D IR spectroscopy for both
inter-molecular\cite{Tanimura-ACR-2009-42} and intramolecular
vibrations.\cite{Ishizaki-JCP-2006-125,
  Ishizaki-JPC-2007-111,Sakurai-JPCA-2011-115} An extension to bridge
between the inter- and intra-molecular modes using an extension of the
present model should also be possible.  We leave such extensions to
future studies, in accordance with progress in experiments and
simulations.

\begin{acknowledgments}
 Financial support from a Grant-in-Aid for Scientific Research
 (A26248005) from the Japan Society for the Promotion of Science is
 acknowledged. T. Ikeda is thankful to T. Hasegawa and A. Sakurai for
 their suggestion to solve the hierarchal Fokker-Planck equations in
 the Hermite representation.
\end{acknowledgments}

\appendix

\section{The Hierarchal Fokker-Planck Equations in the Hermite Representation}
\label{sec:hermite_rep}
In the case of a strong system-bath coupling strength $\zeta $,
Eqs. \eqref{eq:heom1} and \eqref{eq:heom2} converge very slowly as
difference equations with a discrete mesh in the phase space. By
expanding in terms of Hermite functions in the momentum direction,
\cite{Risken-Book} the convergence can be improved. In this case, the
HEOM become simultaneous equations for the coefficients of the
expression. We expand the distribution function as follows:
\begin{align}
  W^{(n)}(q,p;t)&=\psi _{0}e^{-\beta U/2}\sum _{k=0}^{\infty }c_{k}^{(n)}(q;t)\psi _{k}(p),
\end{align}
where $\psi _{k}(p)$ is the $k$th Hermite function,
\begin{align}
  \psi _{k}(p)&=\frac{1}{\sqrt {\mathstrut 2^{k}k!a\sqrt {\mathstrut \pi }}}H_{k}\left(\frac{p}{a}\right)\exp \left(-\frac{p^{2}}{2a^{2}}\right), 
\end{align}
with $H_{k}(x)$ is the $k$th Hermite polynomial and $a=\sqrt {2m /\beta }$.

The Liouvillian of the system (given in Eq. \eqref{eq:liouvillian}) and the
relaxation operators (given in Eqs. \eqref{eq:phi} and \eqref{eq:theta}) are
expressed as
\begin{align}
  \bar{L}&= -b^{-}D^{+} + b^{+}D^{-},\\
  \bar{\Phi }&=V'(q)\sqrt {\mathstrut \frac{\beta }{m}}b^{+},
\end{align}
and
\begin{align}
  \bar{\Theta } &=-\zeta V'(q)\sqrt {\mathstrut \frac{m}{\beta }}b^{-},
\end{align}
where
$\bar{A} \equiv e^{\beta p^{2}/4m+\beta U/2}\hat{A}e^{-\beta p^{2}/4m-\beta U/2}$,
\begin{align}
  b^{\pm } &\equiv \frac{1}{2}\sqrt {\mathstrut \frac{\beta }{m}}p\mp \sqrt {\mathstrut \frac{m}{\beta }}\frac{\partial }{\partial p},
\end{align}
and
\begin{align}
  D^{\pm } &\equiv \frac{1}{2}\sqrt {\mathstrut \frac{\beta }{m}}\frac{\partial U}{\partial q}\mp \frac{1}{m}\sqrt {\mathstrut \frac{m}{\beta }}\frac{\partial }{\partial q}.
\end{align}
Then, the equations of motion for the coefficients $c_{k}^{(n)}(q:t)$ are reduced to
\begin{align}
  \begin{aligned}
    \frac{\partial c_{k}^{(n)}}{\partial t} &= \sqrt {\mathstrut k+1}D^{+}c_{k+1}^{(n)}-\sqrt {\mathstrut k}D^{-}c_{k-1}^{(n)}-n\gamma c_{k}^{(n)}\\
    &\qquad +n\gamma \zeta V'(q)\sqrt {\mathstrut \frac{m}{\beta }}\sqrt {\mathstrut k+1}c_{k+1}^{(n-1)}-V'(q)\sqrt {\mathstrut \frac{\beta }{m}}\sqrt {\mathstrut k}c_{k-1}^{(n+1)}\\
  \end{aligned}
  \label{eq:hermite_rep1}
\end{align}
for $0\leq n<N$ and
\begin{align}
  \begin{aligned}
    \frac{\partial c_{k}^{(N)}}{\partial t} &= \sqrt {\mathstrut k+1}D^{+}c_{k+1}^{(N)}-\sqrt {\mathstrut k}D^{-}c_{k-1}^{(N)}-N\gamma c_{k}^{(N)}\\
    &\qquad +N\gamma \zeta V'(q)\sqrt {\mathstrut \frac{m}{\beta }}\sqrt {\mathstrut k+1}c_{k+1}^{(N-1)}-\zeta V'(q)^{2}kc_{k}^{(N)}.
    \label{eq:hermite_rep2}
  \end{aligned}
\end{align}
To carry out the numerical calculation, we chose $k_{\mathrm{max}}$ so as to satisfy
$c_{k}\approx 0\ (k\geq k_{\mathrm{max}})$ and solved
Eqs. \eqref{eq:hermite_rep1}-\eqref{eq:hermite_rep2} as $k_{\mathrm{max}}$
simultaneous equations.

\section{The RTT component, $\CorRTTp$}
\label{sec:rtt}
\begin{figure}[bthp]
  \centering
  \includegraphics[scale=1.0]{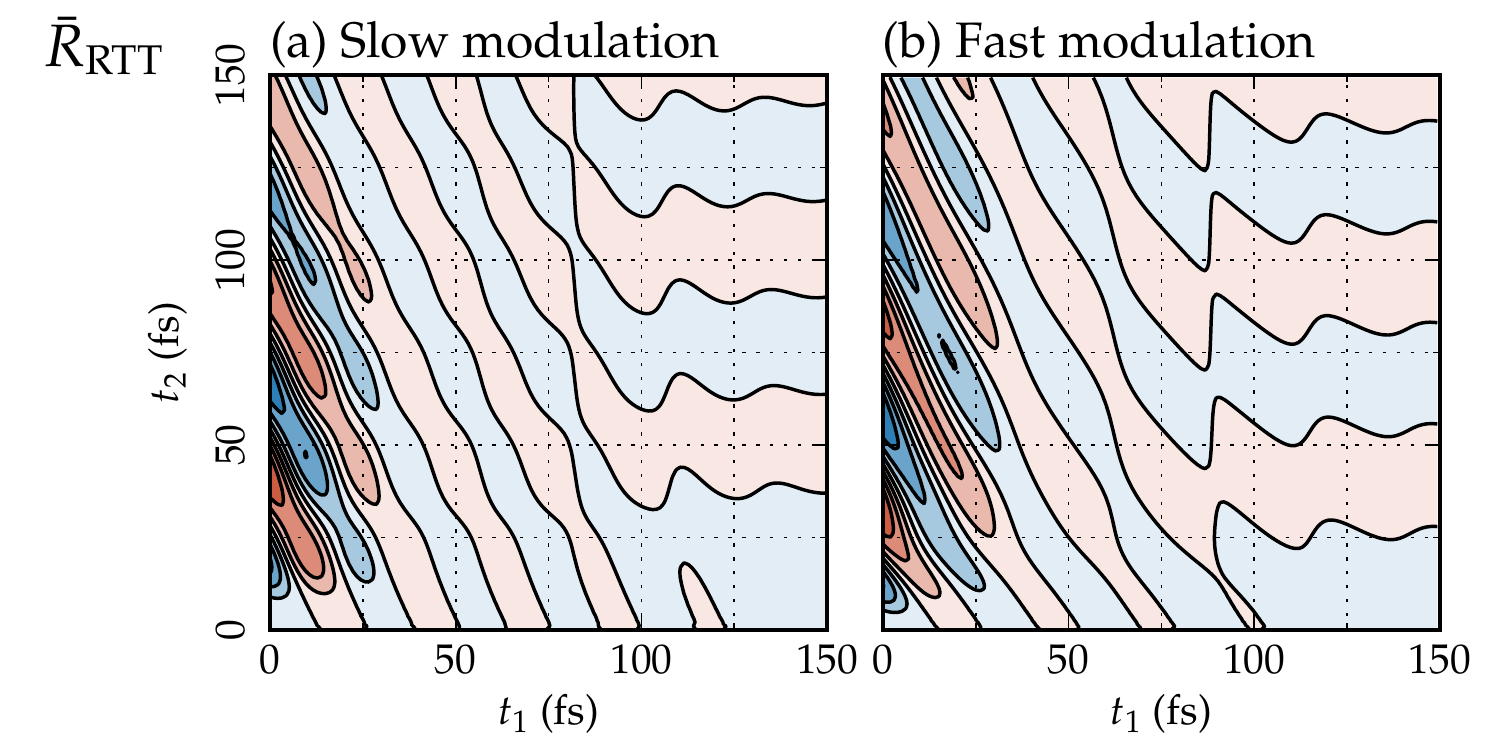}
  \caption{The 2D profiles of the $\CorRTT$ component calculated using
    the harmonic SL BO model ($g_{3}=0$) for (a) the slow modulation case
    ($\zeta = 1.0\ \omega _{0}$ and $\gamma = 0.5\ \omega _{0}$) and (b) the fast
    modulation case ($\zeta = 0.49\ \omega _{0}$ and $\gamma =\infty $). Contours in
    red and blue represent positive and negative values,
    respectively. }
  \label{fig:h_rtt}
\end{figure}
\begin{figure}[bthp]
  \centering
  \includegraphics[scale=1.7]{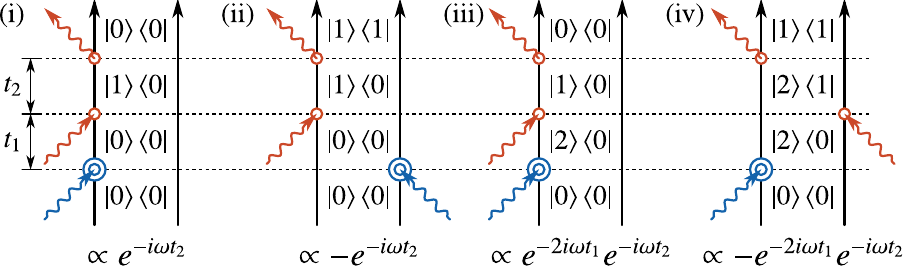}
  \caption{The Liouville paths involved in $\CorRTT$. The red circles
    and the blue double circles represent single and double quantum
    transitions, respectively. The four other paths, which are the
    Hermitian conjugates of the above paths, are not presented. In the
    harmonic case, the propagators from (i) and (ii), and those from
    (iii) and (iv) cancel, and the signal from this component
    vanishes.}
  \label{fig:diagrams_211}
\end{figure}
In Fig. \ref{fig:h_rtt}, we depict the $\CorRTT$ component for the (a)
slow and (b) fast modulation cases, calculated using the harmonic SL BO
model ($g_{3}=0$).  Here, we chose the same parameter values as in IV-A.
The estimated signal intensity for $\CorRTT$ is presented in Table
\ref{tab:demo}.

First we should note that the signal intensity of this component is
very weak.  To illustrate this point, we present the optical Liouville
paths for $\CorRTT$ in Fig. \ref{fig:diagrams_211}.  In the harmonic
case, the diagrams in Figs. \ref{fig:diagrams_211}(i) and
\ref{fig:diagrams_211} (ii), and those in
Figs. \ref{fig:diagrams_211}(iii) and \ref{fig:diagrams_211} (iv)

cancel, respectively, and the signal from $\CorRTT$ vanishes. However,
in the anharmonic case and/or the SL BO case, the diagrams in
Figs. \ref{fig:diagrams_211}(iii) and \ref{fig:diagrams_211} (iv) may
survive, although those in Figs. \ref{fig:diagrams_211}(i) and (ii)
always cancel.  This is because the transition frequencies between
$|0\rangle $ and $|1\rangle $ and between $|1\rangle $ and $|2\rangle $ involved in the $t_{2}$
period of Figs. \ref{fig:diagrams_211}(iii) and \ref{fig:diagrams_211}
(iv) are different in these cases. Thus, we observe the signal
displayed in Fig. \ref{fig:h_rtt} with nodes along the
$2t_{1}+t_{2}=n\pi /\omega $ directions. Its intensity is weak, however,
because of the cancelation.

\section{The 2D profiles of the signal components for different LL+SL couplings}
In this appendix, we present the profiles of the 2D THz-Raman signals
in terms of the (i) AH, (ii) RTT, (iii) TRT and (iv) TTR components to
clarify the roles of the LL and/or SL interactions.  These results can
be used to elucidate the key features of the inter-molecular
interactions as revealed by 2D signals obtained from experiments or
simulations.

To carry out the numerical calculations, we set the frequency,
anharmonicity, and temperature as $\omega _{0}=600$ $\mathrm{cm}^{-1}$ , $g_{3}=0.1$, and
$T=300$ $\mathrm{K}$. Then we calculated the signal components for the LL
($V_{\mathrm{LL}}=1$, $V_{\mathrm{SL}}=0$), SL ($V_{\mathrm{LL}}=0$, $V_{\mathrm{SL}}=1$), and LL+SL ($V_{\mathrm{LL}}=0.5$,
$V_{\mathrm{SL}}=1$) models in the slow ($\gamma = 0.5\ \omega _{0}$) and fast
($\gamma =\infty $) modulation cases. Because the effective coupling strength
depends on $\gamma $,\cite{Tanimura-JCP-accepted} we adjusted $\zeta $ in the
fast and slow modulation cases in order to have the same maximum
intensity in the absorption spectrum.

\begin{figure}[bthp]
  \centering
  \includegraphics[scale=0.8]{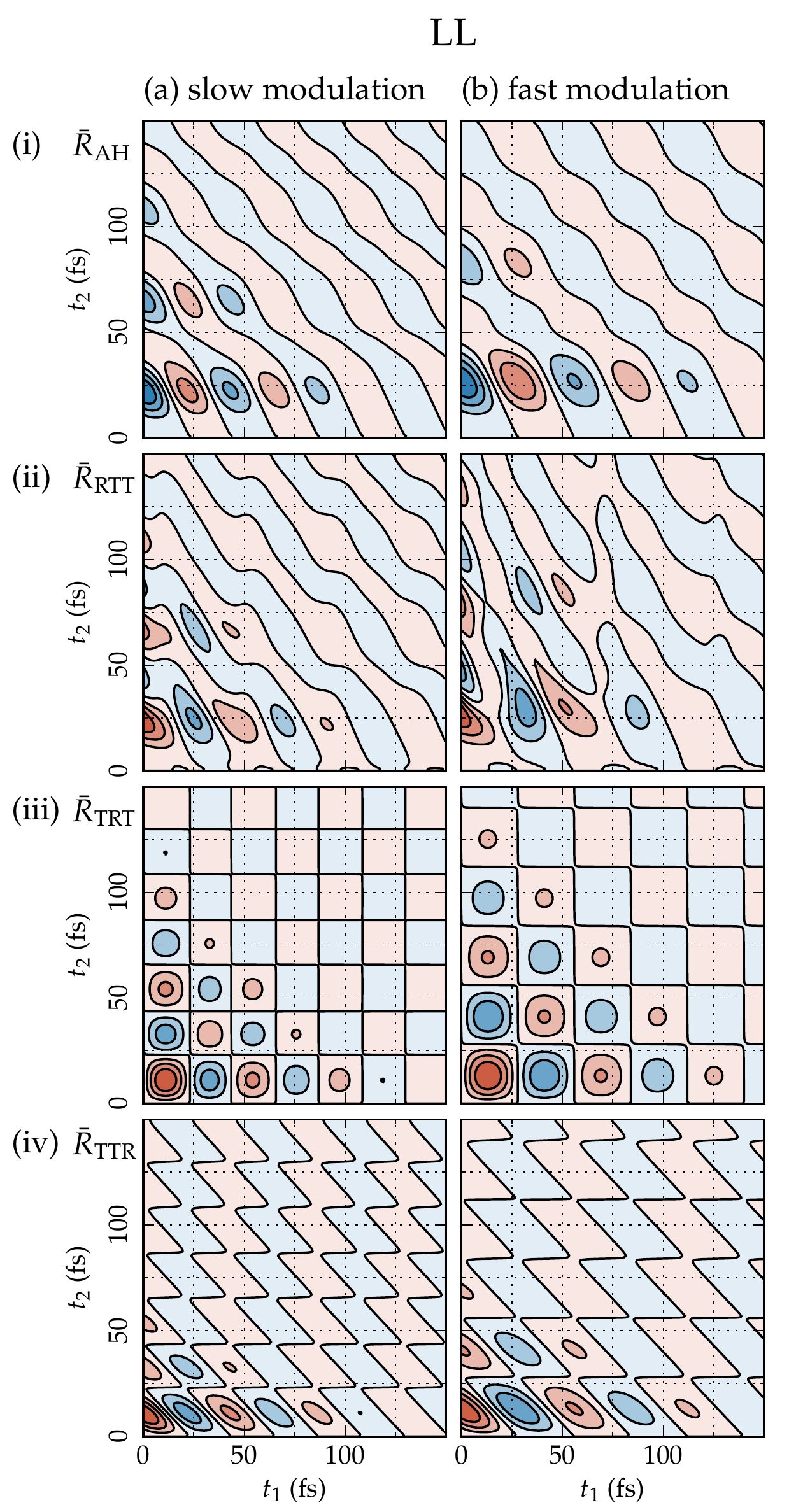}
  \caption{The 2D profiles of the (i) $\CorAH$, (ii) $\CorRTT$, (iii)
    $\CorTRT$, and (iv) $\CorTTR$ components for the (a) slow
    modulation ($\zeta =1.5\ \omega _{0}$ and $\gamma = 0.5\ \omega _{0}$) and fast
    modulation ($\zeta =0.2\ \omega _{0}$ and $\gamma =\infty $) cases obtained from
    the anharmonic ($g_{3}=0.1$) LL BO model.  Contours in red and blue
    represent positive and negative values, respectively. }
  \label{fig:ll}
\end{figure}
\begin{figure}[bthp]
  \centering
  \includegraphics[scale=0.8]{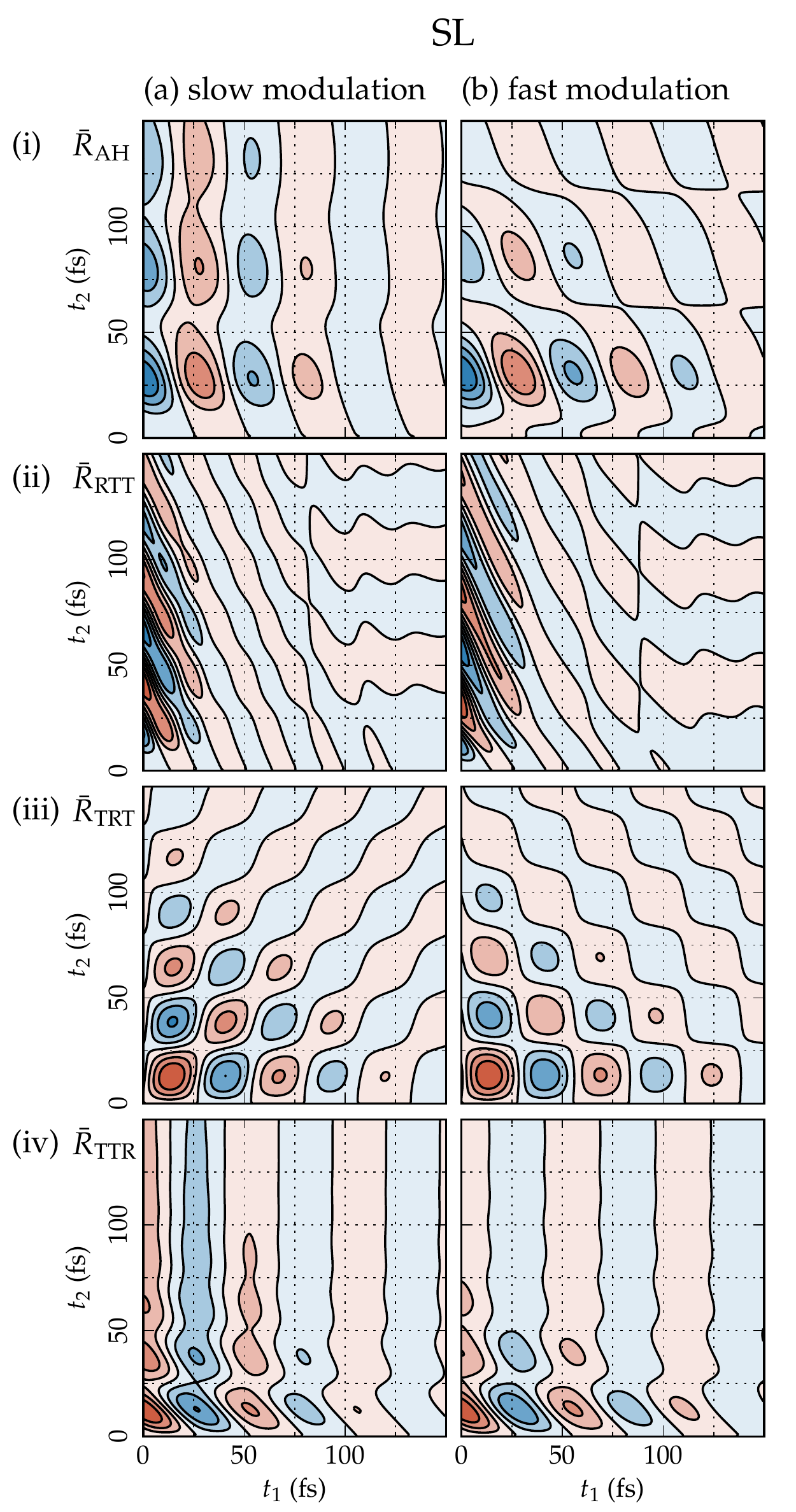}
  \caption{ The 2D profiles of the (i) $\CorAH$, (ii) $\CorRTT$, (iii)
    $\CorTRT$, and (iv) $\CorTTR$ components for the (a) slow ($\zeta =\omega _{0}$
    and $\gamma = 0.5\ \omega _{0} $) and fast modulation ($\zeta =0.49\ \omega _{0}$
    and $\gamma =\infty $) cases obtained from the anharmonic ($g_{3}=0.1$) SL BO
    model.  Contours in red and blue represent positive and negative
    values, respectively.  }
  \label{fig:sl}
\end{figure}
\begin{figure}[bthp]
  \centering
  \includegraphics[scale=0.8]{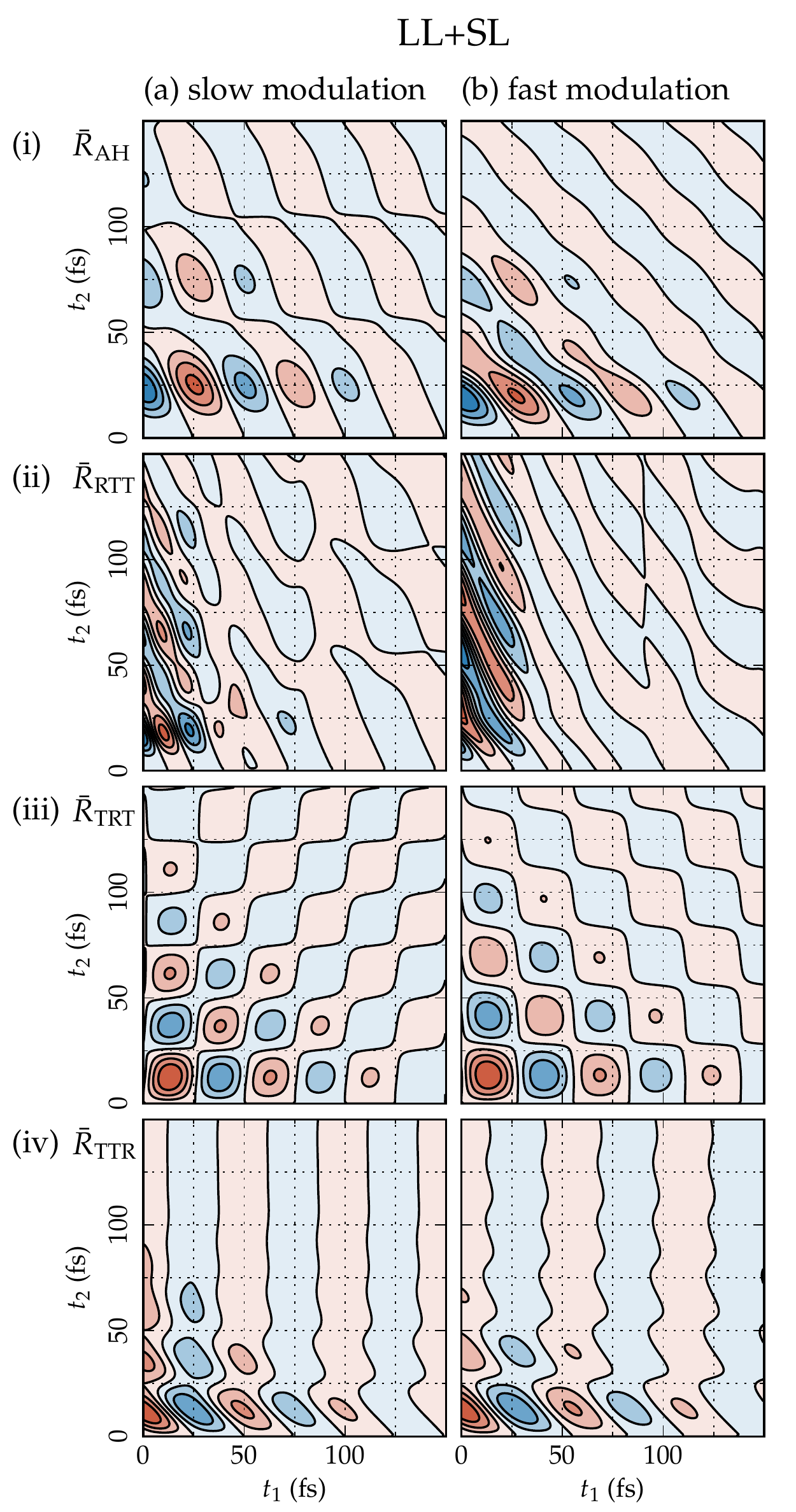}
  \caption{ The 2D profiles of the (i) $\CorAH$, (ii) $\CorRTT$, (iii)
    $\CorTRT$, and (iv) $\CorTTR$ components for the (a) slow
    ($\zeta =1.35\ \omega _{0}$ and $\gamma =0.5\ \omega _{0}$) and fast modulation
    ($\zeta =0.32\ \omega _{0}$ and $\gamma =\infty $) cases obtained from the
    anharmonic ($g_{3}=0.1$) LL+SL BO model ($V_{\mathrm{LL}}=0.5$ and
    $V_{\mathrm{SL}}=1$). Contours in red and blue represent positive and negative
    values, respectively.  }
  \label{fig:ll_sl}
\end{figure}

\begin{table*}[tb]  
  \caption{Relative intensities of the signal components for the LL
    ($V_{\mathrm{LL}}=1$, $V_{\mathrm{SL}}=0$), SL ($V_{\mathrm{LL}}=0$, $V_{\mathrm{SL}}=1$), and LL+SL LL
    ($V_{\mathrm{LL}}=0.5$, $V_{\mathrm{SL}}=1$) anharmonic BO model ($g_3=0.1$). The
    intensities are estimated from the maximum peak values of the
    signals. The obtained intensities were normalized with respect to the
    intensity of $\CorTRT$ in the slow modulation case.\\ }
  \label{tab:intensities}

  \begin{tabular}{cc|ccccc}
    \hline \hline
    Model & Modulation & $\CorAH$ & $\CorRTT$ & $\CorTRT$  & $\CorTTR$ \\
    \hline
    LL    & slow & $0.04$ & $<0.01$ &   $1$  & $1.00$ \\
    LL    & fast & $0.07$ &   $0.01$ & $1.22$ & $1.22$ \\
    SL    & slow & $0.05$ &   $0.23$ & $1.28$ & $1.26$ \\
    SL    & fast & $0.06$ &   $0.14$ & $1.09$ & $1.04$ \\
    LL+SL & slow & $0.37$ &   $0.16$ & $1.15$ & $1.11$ \\
    LL+SL & fast & $0.15$ &   $0.10$ & $1.16$ & $1.13$ \\
    \hline \hline  
  \end{tabular}
\end{table*}

The 2D profiles of the (i) $\CorAH$, (ii) $\CorRTT$, (iii) $\CorTRT$,
and (iv) $\CorTTR$ components for the LL, SL and LL+SL models are
presented in Figs. \ref{fig:ll}-\ref{fig:ll_sl}, respectively. The
estimated signal intensities for each component are listed in Table
\ref{tab:intensities}.

Because the anharmonicity is not strong, the calculated results
appearing in Figs. \ref{fig:ll}(iii) and \ref{fig:ll}(iv) are similar
to those predicted by Eqs. \eqref{eq:cor_trtHBO} and
\eqref{eq:cor_ttrHBO}, respectively. This indicates that the node
lines in $\CorTRT$ correspond to $t_{1}= 2n\pi /\omega $ and $t_{2}= 2m\pi /\omega $, whereas
those in $\CorTTR$ correspond to $t_{1}+2t_{2}=2n\pi /\omega $, where $n$ and $m$ are
any integers. Note that, due to the effect of the heat bath, the
fundamental frequency $\omega $ is shifted from
$\omega _{0}$.\cite{Tanimura-JCP-accepted} The components $\CorAH$ and $\CorRTT$
in the LL case appear solely due to the anharmonicity, while those
presented in Figs. \ref{fig:ah_ah} and \ref{fig:h_rtt} arise from both
the SL interaction and the anharmonicity in the former case and just
the SL interaction in the latter case. The profiles in the fast
modulation limit of the SL mode presented in Figs. \ref{fig:ah_ah}(b)
and \ref{fig:sl}(b) are similar to those for the anharmonic LL model
presented in Fig. \ref{fig:ah_ah}, because this limit corresponds to
the case of homogeneously distributed oscillators, illustrated in
Fig. \ref{fig:homo_inhomo}(a). The intensity of the $\CorRTT$
component in the LL case is significantly weaker than those of the
other components, because of the cancelation explained in Appendix
B. As in the case of Fig. \ref{fig:h_rtt}, we observe nodes along the
$2t_{1}+t_{2}=2n\pi /\omega $ direction, although they are not clear.

Although we included anharmonicity in the SL case depicted in
Fig. \ref{fig:sl}, the overall profiles of the signals are similar to
those in the harmonic case, presented in Figs. \ref{fig:h_trt},
\ref{fig:h_ttr} and \ref{fig:h_rtt}. This indicates that the effects
of the SL interactions are dominant in these profiles. This is also
true in the LL+SL case considered in Fig. \ref{fig:ll_sl}, as
indicated by the similarity of the profiles there with those in
Figs. \ref{fig:h_trt}, \ref{fig:h_ttr}, \ref{fig:h_rtt} and
\ref{fig:sl}. The intensity of the $\CorAH$ component in the LL+SL
case, however, is much larger than that in the other cases, because
the LL+SL interaction induces transitions through the cubic
interactions $V_{\mathrm{LL}}V_{\mathrm{SL}}\hat{q}(t'')\hat{q}^{2}(t')$ and $V_{\mathrm{LL}}V_{\mathrm{SL}}\hat{q}^{2}(t'')\hat{q}(t')$,
derived from $V(\hat{q}(t''))V(\hat{q}(t'))$. These interactions can induce a
signal in the $\CorAH$ component, even in the harmonic case, involving
only Gaussian integrals in the response function, for example, as
$ V_{\mathrm{LL}}V_{\mathrm{SL}}\mathrm{Tr}\{q(t_{1}+t_{2})\hat{q}(t'')\hat{q}(t_{1})\hat{q}^{2}(t')\hat{q}(0)\exp (-\beta \hat{H}_{\mathrm{S}})\}$.\cite{Kato-JCP-2004-120}
Then, the profile of the $\CorAH$ component in the LL+SL case exhibits
some similarity to that in the LL model, presented in
Fig. \ref{fig:ll}(iii), because the dipole and linear polarizability
operators associated with the system-bath interactions, for example,
$ \hat{q}(t'')\hat{q}(t_{1})\hat{q}^{2}(t')$ and $ \hat{q}(t_{1}+t_{2})\hat{q}(t'')\hat{q}^{2}(t')$, induce the
same DQ and ZQ transitions as nonlinear polarizability.




\clearpage
\begin{thebibliography}{99}\label{sec:TeXbooks}
\bibitem{Mukamel-Book}
  S. Mukamel, \textit{Principles of Nonlinear Optical Spectroscopy} (Oxford University Press, Oxford, 1995).
\bibitem{Cundiff-PT-2013-7}
  S. T. Cundiff and S. Mukamel, Physics Today \textbf{7}, 44 (2013).
\bibitem{Mukamel-ACR-2009-42}
  S. Mukamel, Y. Tanimura, and P. Hamm, Acc. Chem. Res. \textbf{42}, 1207 (2009).

\bibitem{Asbury-JPCA-2004-108}
  J. B. Asbury, T. Steinel, C. Stromberg, S. A. Corcelli, C. P. Lawrence, J. L. Skinner, and M. D. Fayer, J. Phys. Chem. A \textbf{108}, 1107 (2004).
\bibitem{Auer-PNAS-2007-104}
  B. Auer, R. Kumar, J. R. Schmidt, and J. L. Skinner, Proc. Nat. Acad. Sci. \textbf{104}, 14215 (2007).

\bibitem{Tanimura-JCP-1993-99}
  Y. Tanimura and S. Mukamel, J. Chem. Phys. \textbf{99}, 9496 (1993).
 
\bibitem{Kaufman-PRL-2002-88} 
  L. J. Kaufman, J. Heo, L. D. Ziegler, and G. R. Fleming, Phys. Rev. Lett. \textbf{88}, 207402 (2002).
\bibitem{Kubarych-IRPC-2003-22} 
  K. J. Kubarych, C. J. Milne, and R. J. D. Miller, Int. Rev. Phys. Chem. \textbf{22}, 497 (2003).
\bibitem{Kubarych-JCP-2002-116} 
  K. J. Kubarych, C. L. Milne, S. Lin, V. Astinov, and R. J. D. Miller, J. Chem. Phys. \textbf{116}, 2016 (2002).
\bibitem{Kubarych-CPL-2003-369} 
  K. J. Kubarych, C. L. Milne, and R. J. D. Miller, Chem. Phys. Lett. \textbf{369}, 635 (2003).
\bibitem{Milne-JPCB-2006-40} 
  C. L. Milne, Y. L. Li, T. L. C. Jansen, L. Huang, and J. D. Miller, J. Phys. Chem. B \textbf{110}, 19867 (2006).
\bibitem{Li-JCP-2008-128} 
  Y. L. Li,  L. Huang, R. J. D. Miller, T. Hasegawa, and Y. Tanimura, J. Chem. Phys. \textbf{128}, 234507 (2008).

\bibitem{Hamm-JCP-2012-136} 
  P. Hamm and J. Savolainen, J. Chem. Phys. \textbf{136}, 094516 (2012).
\bibitem{Hamm-JCP-2012-136-note} 
  P. Hamm, J. Savolainen, J. Ono, and Y. Tanimura, J. Chem. Phys. \textbf{136}, 236101 (2012).
\bibitem{Ito-JCP-2014-141} 
  H. Ito, T. Hasegawa, and Y. Tanimura, J. Chem. Phys. \textbf{141}, 124503 (2014).
\bibitem{Hamm-JCP-2014-141}
P. Hamm, J. Chem. Phys. \textbf{141}, 184201 (2012).
\bibitem{Savolainen-PNAS-2013-110} 
  J. Savolainen, S. Ahmed, and P. Hamm, Proc. Natl. Acad. Sci. U.S.A. \textbf{110}, 20402 (2013).

\bibitem{Berens-JCP-1981-74} %
  P. H. Berens and K. R. Wilson, J. Chem. Phys. \textbf{74}, 4872 (1981).
  
\bibitem{Ma-PRL-2000-85} 
  A. Ma and R. M. Stratt, Phys. Rev. Lett. \textbf{85}, 1004 (2000).
\bibitem{Saito-PRL-2002-88} 
  S. Saito and I. Ohmine, Phys. Rev. Lett. \textbf{88}, 207401 (2002).
\bibitem{Saito-JCP-2003-119} 
  S. Saito and I. Ohmine, J. Chem. Phys. \textbf{119}, 9073 (2003).
\bibitem{Saito-JCP-2006-125} 
  S. Saito and I. Ohmine, J. Chem. Phys. \textbf{125}, 084506 (2006).
\bibitem{Nagata-JCP-2007-126} 
  Y. Nagata, Y. Tanimura, and S. Mukamel, J. Chem. Phys. \textbf{126}, 204703 (2007).

\bibitem{Jansen-JCP-2001-114}
  T. I. C. Jansen, J. G. Snijders, and K. Duppen, J. Chem. Phys. \textbf{114}, 10910 (2001).
\bibitem{Jansen-JCP-2003-63}
  T. I. C. Jansen, K. Duppen, and J. G. Snijders, Phys. Rev. B \textbf{67}, 134206 (2003).
  
\bibitem{Hasegawa-JCP-2006-125} 
  T. Hasegawa and Y. Tanimura, J. Chem. Phys. \textbf{125}, 074512 (2006).

\bibitem{Tanimura-ACR-2009-42}  
  Y. Tanimura and A. Ishizaki, Acc. Chem. Res. \textbf{42}, 1270 (2009).

\bibitem{Ishizaki-JCP-2006-125} 
  A. Ishizaki and Y. Tanimura, J. Chem. Phys. \textbf{125}, 084501 (2006).
\bibitem{Ishizaki-JPC-2007-111} 
  A. Ishizaki and Y. Tanimura, J. Phys. Chem.\textbf{111}, 9269 (2007).

\bibitem{Tanimura-JPSJ-1989-58} 
  Y. Tanimura and R. Kubo, J. Phys. Soc. Jpn. \textbf{58}, 101 (1989).
\bibitem{Tanimura-PRA-1991-43}
  Y. Tanimura and P. G. Wolynes, Phys. Rev. A \textbf{43}, 4131 (1991).
\bibitem{Tanimura-JCP-1992-96}
  Y. Tanimura and P. G. Wolynes, J. Chem. Phys. \textbf{96}, 8485 (1992).
\bibitem{Steffen-JPSJ-2000-69}  
  T. Steffen and Y. Tanimura, J. Phys. Soc. Jpn. \textbf{69}, 3115 (2000).
\bibitem{Tanimura-JPSJ-2000-69} 
  Y. Tanimura and T. Steffen, J. Phys. Soc. Jpn. \textbf{69}, 4095 (2000).
\bibitem{Kato-JCP-2002-117}
  T. Kato and Y. Tanimura, J. Chem. Phys. \textbf{117}, 6221 (2002).
\bibitem{Kato-JCP-2004-120}
  T. Kato and Y. Tanimura, J. Chem. Phys. \textbf{120}, 260 (2004).
\bibitem{Tanimura-JPSJ-2006-75} 
  Y. Tanimura, J. Phys. Soc. Jpn. \textbf{75}, 082001 (2006).
\bibitem{Tanimura-JCP-accepted}
  Y. Tanimura, \textit{Real-Time and Imaginary-Time Quantum Hierarchal Fokker-Planck Equations,} J. Chem. Phys. \textit{accepted}.

\bibitem{Sagarik-JCP-1987-86}   
  K. P. Sagarik and R. Ahlrichs, J. Chem. Phys. \textbf{86}, 5117 (1987).
\bibitem{Wojcik-JCP-2000-113}   
  M. C. Wojcik, K. Hermansson, and H. O. G. Siegbahn, J. Chem. Phys. \textbf{113}, 374 (2000).
\bibitem{Abascal-JCP-2005-23}   
  J. L. F. Abascal and C. Vega, J. Chem. Phys. \textbf{123}, 234505 (2005).
\bibitem{Walser-JCP-2000-112}   
  R. Walser, A. E. Mark, W. F. van Gunsteren, M. Lauterbach, and G. Wipff, J. Chem. Phys. \textbf{112}, 10450 (2000).

\bibitem{Applequist-JACS-1972-94} 
  J. Applequist, J. R. Carl, and K.-K. Fung, J. Am. Chem. Soc. \textbf{94}, 2952 (1972).
\bibitem{Huiszoon-MP-1986-58}   
  C. Huiszoon, Mol. Phys. \textbf{58}, 865 (1986).
\bibitem{Ladanyi-CPL-1985-121} 
  B. M. Ladanyi, Chem. Phys. Lett. \textbf{121}, 351 (1985).

\bibitem{Hasegawa-JCP-2008-128}
  T. Hasegawa and Y. Tanimura, J. Chem. Phys. \textbf{128}, 064511 (2008).
\bibitem{Yagasaki-JCP-2008-128}
  T. Yagasaki and S. Saito, J. Chem. Phys. \textbf{128},  154521 (2008).
\bibitem{Jeon-JPB-2014-118}
  J. Jeon and M. Cho, J. Phys.B, \textbf{118}, 8184  (2014).

\bibitem{Sakurai-JPCA-2011-115}  
  A. Sakurai and Y. Tanimura, J. Phys. Chem. A \textbf{115}, 4009 (2011).

\bibitem{Caldeira-PA-1983-121}  
  A. O. Caldeira and A. J. Leggett, Physica A \textbf{121}, 587 (1983).
\bibitem{Grabert-PR-1988-115}   
  H. Grabert, P. Schramm, and G. L. Ingold, Phys. Rep. \textbf{168}, 115 (1988).
\bibitem{Okumura-PRE-1997-56}   
  K. Okumura and Y. Tanimura, Phys. Rev. E. \textbf{56}, 2747 (1997).  

\bibitem{Saito-JCP-108-1998}
  S. Saito and I. Ohmine, J. Chem. Phys. \textbf{108}, 240 (1998).


\bibitem{Risken-Book}           
  H. Risken, \textit{The Fokker-Planck Equation, 2nd ed.} (Springer, Berlin, 1989).
  
\bibitem{Tanimura-CP-1998-233}
  Y. Tanimura, Chem. Phys. \textbf{233}, 217 (1998).

\bibitem{Okumura-JCP-1996-105}  
  K. Okumura and Y. Tanimura, J. Chem. Phys. \textbf{105}, 7294 (1996).
\bibitem{Okumura-CPL-1997-277}  
  K. Okumura, and Y. Tanimura, Chem. Phys. Lett. \textbf{277}, 159. (1997).


  
\bibitem{Bosma-JCP-1993-98}     
  W. B. Bosma, L. E. Fried, and S. Mukamel, J. Chem. Phys. \textbf{98}, 4413 (1993).
\bibitem{Larsen-JCP-2006-125}   
  R. W. Larsen and M. A. Suhm, J. Chem. Phys. \textbf{125}, 154314 (2006).
  
\end{thebibliography}

\end{document}